\documentclass[11pt]{article}

\usepackage{amssymb,amsmath,amsfonts,eurosym,geometry,ulem,graphicx,caption,color,setspace,sectsty,comment,footmisc,caption,natbib,pdflscape,array,natbib,har2nat,booktabs,threeparttable,array,siunitx}

\usepackage[hidelinks]{hyperref}
\usepackage{soul}
\usepackage{tabularx}
\newcolumntype{L}[1]{>{\raggedright\let\newline\\arraybackslash\hspace{0pt}}m{#1}}
\newcolumntype{C}[1]{>{\centering\let\newline\\arraybackslash\hspace{0pt}}m{#1}}
\newcolumntype{R}[1]{>{\raggedleft\let\newline\\arraybackslash\hspace{0pt}}m{#1}}

\usepackage[utf8]{inputenc} % allow utf-8 input
\usepackage[T1]{fontenc}    % use 8-bit T1 fonts
\usepackage{hyperref}       % hyperlinks
\usepackage{url}            % simple URL typesetting
\usepackage{booktabs}       % professional-quality tables
\usepackage{xcolor}
\usepackage{amsfonts}       % blackboard math symbols
\usepackage{tabularx, makecell}
\usepackage{nicefrac}       % compact symbols for 1/2, etc.
\usepackage{microtype}      % microtypography
\usepackage{lipsum}
\usepackage{fancyhdr}       % header
\usepackage{graphicx}       % graphics
\graphicspath{{media/}}     % organize your images and other figures under media/ folder
\usepackage{graphicx, enumitem,subfig}
\usepackage{adjustbox}
\usepackage{wrapfig}
\usepackage{setspace}
\usepackage{subfig}
\usepackage{array}
\begin{document}
\newgeometry{left=0.8in, right=0.8in, top=0.7in, bottom=0.8in}
\begin{titlepage}

\title{The Adoption and Efficacy of Large Language Models: \\ Evidence From Consumer Complaints in the Financial Industry}
% \title{\Large{\textbf{The Temporal Pattern and Impact of Large Language Model Adoption: \\Evidence from Consumer Complaints in the Financial Industry}}}
% \title{\Large{\textbf{\textcolor{red}{The Temporal Pattern and Impact of Large Language Model Adoption: \\Evidence from Consumer Complaints in the Financial Industry}}}}

\author{
  Minkyu Shin\thanks{Minkyu Shin: Assistant Professor of Marketing, City University of Hong Kong, minkshin@cityu.edu.hk. $^{\dagger}$Jin Kim: Postdoctoral Research Associate, Northeastern University, jin.kim1@northeastern.edu. $^{\ddagger}$Jiwoong Shin: Professor of Marketing, Yale School of Management, jiwoong.shin@yale.edu.  } \and Jin Kim$^{\dagger}$ \and Jiwoong Shin$^{\ddagger}$
}

\date{\today}
\maketitle

\begin{abstract}

\noindent Large Language Models (LLMs) are reshaping consumer decision-making, particularly in communication with firms, yet our understanding of their impact remains limited. This research explores the effect of LLMs on consumer complaints submitted to the Consumer Financial Protection Bureau from 2015 to 2024, documenting the adoption of LLMs for drafting complaints and evaluating the likelihood of obtaining relief from financial firms. We analyzed over 1 million complaints and identified a significant increase in LLM usage following the release of ChatGPT. We find that LLM usage is associated with an increased likelihood of obtaining relief from financial firms. To investigate this relationship, we employ an instrumental variable approach to mitigate endogeneity concerns around LLM adoption. Although instrumental variables suggest a potential causal link, they cannot fully capture all unobserved heterogeneity. To further establish this causal relationship, we conducted controlled experiments, which support that LLMs can enhance the clarity and persuasiveness of consumer narratives, thereby increasing the likelihood of obtaining relief. Our findings suggest that facilitating access to LLMs can help firms better understand consumer concerns and level the playing field among consumers. This underscores the importance of policies promoting technological accessibility, enabling all consumers to effectively voice their concerns.

\vspace{0.3cm}

% \noindent \textbf{Significance Statement:}
% Despite the growing interest in generative AI, most existing research on LLM adoption relies on controlled lab studies or surveys rather than real-world data. This study uniquely leverages large-scale, publicly available datasets—over 1 million consumer complaints submitted to the Consumer Financial Protection Bureau between 2014 and 2024—to rigorously document the timing, heterogeneity, and impact of LLM adoption in actual consumer-firm interactions. We find that the release of ChatGPT spurred a pronounced increase in AI-generated complaint narratives, improving the likelihood of securing relief. By systematically capturing both adoption patterns \textit{and} real-world efficacy, this work represents one of the earliest large-scale empirical investigations into LLM usage. It also provides an accessible testbed for future inquiries, offering critical insights for policymakers, firms, and researchers seeking to understand and shape the rapidly evolving role of AI in consumer advocacy.
% Consumer complaints refined by ChatGPT
% for improved linguistic features were regarded as more likely to receive relief offers
% than the original consumer complaints, demonstrating the LLM’s ability to enhance
% complaint persuasiveness in human communication. More importantly, this tendency
% persisted even when complaints improved for a particular linguistic feature (e.g., clarity)
% were evaluated by human handlers with a preference for a different linguistic feature
% (e.g., professionalism) in their compensation decisions. 
\vspace{0.3cm}
\begin{tabbing}
\noindent\textbf{Competing Interests Statement:} The authors declare no competing interests.  \\
    
    \noindent\textbf{Classification:} Social Science / Psychological and Cognitive Sciences \\

\hspace{1.9cm} Financial Literacy; The Impact of AI 
\end{tabbing}

\end{abstract}
\setcounter{page}{0}
\thispagestyle{empty}

\end{titlepage}
\restoregeometry
\pagebreak \newpage

\doublespacing

\section{Introduction}

\noindent Large Language Models (LLMs) and generative AI are profoundly influencing everyday human activities across a broad spectrum, particularly by enhancing the precision and efficacy of language in human interactions. This impact stems from the fundamental nature of LLMs, which are centered around language—a crucial element in human communication. Tools like ChatGPT are revolutionizing persuasive dialogue by aiding users in drafting and refining messages. This assistance enhances the clarity and persuasiveness of communication, which is crucial for effective interaction. For instance, recent studies show that academic researchers utilize LLMs to improve the clarity and impact of their writing, increasing reader engagement and comprehension, particularly when writing skills do not match the depth of their insights \cite{liang2024mapping}. Turning to the business context, these capabilities of LLMs are particularly relevant in consumer-initiated communications, such as complaints, where the clarity and persuasiveness of consumer messaging are critical for resolving disputes and securing favorable outcomes from firms. In light of this, our research examines the role of LLMs, such as ChatGPT, in enhancing consumer-firm communications. We focus on two main aspects: the adoption of LLMs by consumers, particularly whether their usage in communications is increasing, and the efficacy of these models in enhancing the persuasiveness of communications, thereby potentially improving outcomes for consumers.
% Given this context, the degree to which consumers adopt LLMs for interacting with firms, as well as the actual effectiveness of these interactions, both remain significant empirical questions. 

Evidence of LLM adoption among consumers varies widely according to different surveys. For example, a large-scale survey conducted between November 2023 and January 2024 indicated that ChatGPT is widely utilized across diverse occupational groups in Denmark, achieving adoption rates as high as 65\% among marketing specialists and journalists \cite{humlum2025unequal}. In contrast, a survey by the Pew Research Center in July 2023 reported a comparatively lower adoption rate of 18\% among Americans, which only marginally increased to 23\% by February 2024 \cite{park2023chatgpt, mcclain2024chatgpt}. Notably, most existing studies on LLM adoption rely primarily on survey data \cite{NBERw32966}. Our research contributes as one of the first empirical studies to document adoption patterns using a dataset of more than 1 million consumer complaints spanning nearly a decade, from March 2015 to March 2024. Moreover, by leveraging ZIP code information in the dataset, we trace the trajectory of these adoption patterns and examine their heterogeneity across various sociodemographic groups.

Furthermore, the effectiveness of consumers using LLMs to obtain better responses from firms has yet to be empirically evaluated. While LLMs can enhance the effectiveness of communications by generating content that is coherent, informative, and formal \cite{wu2022autoformalization}, they might also diminish effectiveness by producing verbose or redundant content \cite{saito2023verbosity}, or by failing to capture the subtle nuances of human emotion, which are essential for effective interpersonal communication \cite{shahriar2023let}. 
Additionally, the effective use of LLMs requires substantial domain expertise. Crafting precise prompts often demands significant experience and a thorough understanding of the specific communication context, while assessing the output also necessitates adequate knowledge of the relevant domain. For example, in the finance industry, substantial evidence of a financial illiteracy problem suggests that many consumers may lack the necessary domain knowledge required for effectively leveraging LLMs \cite{lusardi2014economic,fernandes2014financial}. This deficiency indicates that, even if LLMs can enhance the clarity or linguistic sophistication of consumer messages, a lack of domain knowledge may prevent consumers from using them effectively to persuade financial firms and secure favorable outcomes. Therefore, \textit{whether} and \textit{to what extent} LLMs impact the actual outcomes in such scenarios remains an open question.

To investigate the adoption and efficacy of using LLMs in consumer communication, we analyzed the consumer complaint database from the Consumer Financial Protection Bureau (CFPB). This dataset serves as an ideal testbed, providing a large-scale, historical, and detailed view of two-sided consumer-firm communication.  Our data spans nearly a decade from 2015 to 2024, including the pivotal release of ChatGPT on November 30, 2022. This allows us to analyze how both consumer complaints and firm responses change before and after the introduction of the most popular LLM.  The dataset includes over 1.1 million complaints, each containing rich details of the consumer's issue and the firm's response, including the relief decisions made by firms.  We also leverage ZIP code information for each complaint to merge data from the American Community Survey (ACS) database, enabling us to study the role of sociodemographic factors. 

The consumer finance domain, with its inherent communication challenges, is well-suited for examining the impact of LLMs on consumer-firm communications. Financial illiteracy \cite{lusardi2014economic}, along with the complexity of industry jargon and regulatory language, poses significant challenges for consumers~\textemdash~particularly those with limited English proficiency\textemdash~in articulating their concerns effectively. This raises the timely question of whether AI tools like LLMs can assist vulnerable consumers in overcoming these challenges \cite{hermann2023deploying}. This scenario, where consumers are motivated by significant personal stakes to enhance their communications, offers an opportunity to evaluate the potential of LLMs to improve effectiveness in consumer-firm communications. 

In our analysis of the Consumer Financial Protection Bureau's extensive dataset of consumer complaints, we document empirical patterns of LLM adoption and examine variations across sociodemographic groups. By leveraging this variation as an instrumental variable, we find evidence that LLM usage can increase the likelihood of obtaining relief. 
% However, instrumental variables, though suggesting a potential causal link, do not capture all unobserved heterogeneity. To further validate this causal relationship,
However, while instrumental variables indicate a potential causal link, they cannot fully account for all unobserved heterogeneity. To establish this causal relationship directly, 
% To support a causal interpretation directly, 
we conduct controlled laboratory experiments, which confirm that LLM usage to refine and polish consumer complaints significantly increases the probability of receiving a favorable response. These improvements occur without requiring consumers to have domain-specific expertise or a deep understanding of the recipient’s preferences, as LLMs autonomously optimize key linguistic features \cite{niven2019probing,manning2020emergent}. Our experiments demonstrate that this process simultaneously enhances clarity, coherence, fluency, and formality \cite{liu2019s, chen2022effects}, suggesting that LLMs can accommodate the heterogeneous preferences of complaint handlers and increase the likelihood of favorable responses in line with communication accommodation theory \cite{giles1979accommodation}. This comprehensive approach, incorporating both observational and experimental studies, demonstrates the practical benefits of LLMs in enhancing consumer advocacy through clearer and more polished messages.

\section{Analysis of LLM Adoption and Efficacy Using the CFPB Dataset}

\subsection{Datasets \& AI Detection Tool} 
\paragraph{CFPB dataset} The CFPB complaint dataset provides granular information on consumer grievances related to a wide spectrum of financial products and services. A key feature of this dataset is the inclusion of detailed consumer narratives, where individuals describe their experiences in their own words. Unlike informal and often brief reviews on platforms such as Yelp or Amazon, complaints submitted to the CFPB are processed through a formal government channel, positioning them as official documents with the potential to trigger regulatory action or investigations. Given this regulatory context, consumers are likely to craft their complaints with greater care and precision, as these submissions can carry serious implications for firms. This stands in stark contrast to the opinion-oriented, casual nature of reviews on commercial platforms. Moreover, the CFPB’s regulatory framework mandates that firms respond to complaints, offering consumers a clear path to potential redress, which may include refunds, corrections, or other forms of relief based on the outcome of the resolution process.

Beyond the consumer narratives, the dataset includes a comprehensive set of structured variables that provide additional insights into the nature and resolution of the complaints. Each complaint is systematically categorized by issue type (e.g., ``Improper use of your report'' or ``Incorrect information on your report'') and product type (e.g., ``Debt collection'' or ``Credit reporting''). The dataset also records firm responses, categorized as ``Closed with monetary relief,'' ``Closed with non-monetary relief,'' or ``Closed with explanation,'' offering a clear view of how firms address consumer grievances. In our analysis, we define a successful complaint outcome as one that results in either monetary or non-monetary relief, consistent with prior literature \cite{dou2023learning}.\footnote{This definition is particularly relevant in the financial context where distinctions between monetary and non-monetary relief are nuanced. Non-monetary relief, although not immediately quantifiable as direct financial compensation, can yield substantial economic benefits. For example, offering a foreclosure alternative like loan modification or forbearance can prevent home loss and the associated displacement costs—providing significant economic value to the consumer without direct financial payment.} Detailed descriptions of all variables are provided in the \textcolor{blue}{\textit{SI Appendix}, section \ref{si_subsection:data_cfpb}}. Moreover, complaints are geocoded by ZIP code level, allowing for the identification of regional patterns in consumer dissatisfaction across the U.S. The ZIP code data further enables the integration of sociodemographic information from the ACS, facilitating a more granular analysis of how LLM adoption varies across different demographic and socioeconomic groups.

\paragraph{ACS dataset} We used the 2017–2021 ACS 5-year estimates to collect representative sociodemographic information at the ZIP code level in order to examine the heterogeneous adoption of LLMs following the release of ChatGPT in November 2022. For each ZIP code, we included key variables such as employment rate, education level, median income, total number of households, English proficiency, and Internet access. These variables were chosen to capture cross-sectional variation in sociodemographic characteristics across different regions. While a detailed explanation of each variable's calculation is provided in the \textcolor{blue}{\textit{SI Appendix}, section \ref{si_subsection:data_acs}}, we essentially followed the Census Bureau's methodology for generating regional profiles.

\paragraph{AI Detection Tool} To identify complaints likely written with the assistance of LLMs, we employ the paid premium version of Winston AI, one of the leading commercial AI detection tools, to compute an \textit{AI Score} for each complaint in the CFPB dataset.\footnote{It is noteworthy that commercialized, paid versions of AI detection tools exhibit substantially superior performance compared to many free alternatives such as GPTZero \cite{ma2023beyond}. To assess the robustness of our detection results, we cross-checked the results of this detection program with another leading paid AI detection tool, Originality.AI. We observe a high degree of agreement between the two AI detection tools, with an agreement rate of 94.1\%.  These results are provided in the \textcolor{blue}{\textit{SI Appendix}, section \ref{si_subsection:ai_detection_tool}}.} Winston AI calculates a ``Human Score,'' which estimates the probability that a given text was written by a human, as opposed to being generated by LLMs such as ChatGPT, GPT-4, Claude, or Gemini \cite{prillaman2023detector}. For the purposes of our analysis, we define the \textit{AI Score} as `$100 - \text{Human Score}$', representing the estimated probability that a complaint was generated with AI assistance. Further details on the tool's methodology and validity, including the results of its accuracy tests, are provided in the \textcolor{blue}{\textit{SI Appendix}, section \ref{si_subsection:ai_detection_tool}}. In the subsequent analysis, we use the AI Score to examine the prevalence of complaints likely written with LLM assistance and to analyze the relationship between LLM usage and complaint outcomes.

\subsection{Consumer Adoption of LLMs for Writing Complaints in Financial Industry} 
\begin{figure}[t]
    \centering
    \caption{LLM Adoption Observed in the CFPB dataset}
     \includegraphics[width=1\linewidth]{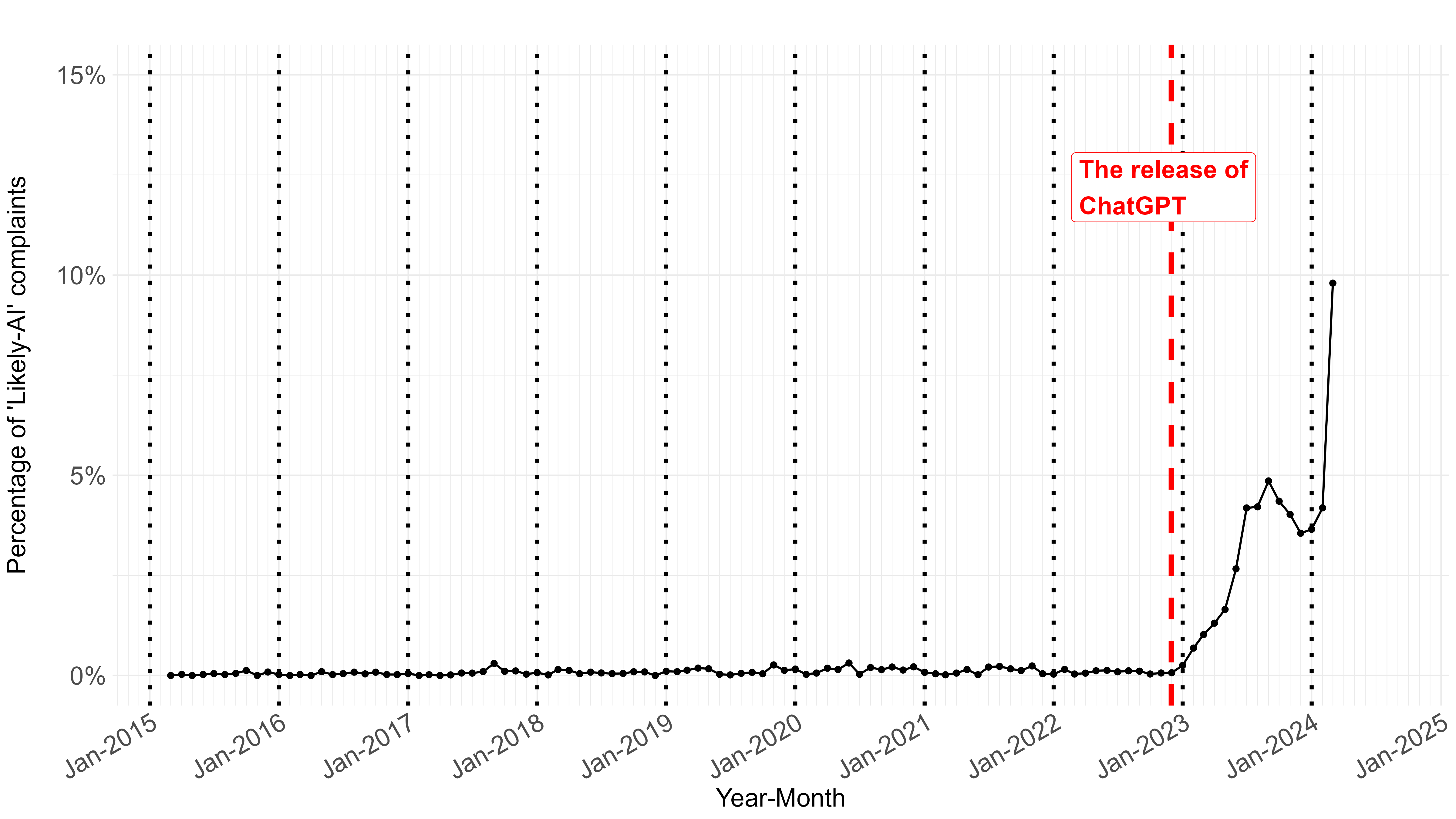}
    \begin{minipage}{0.95\linewidth}
    \vspace{0.5em}
        \small \textit{Note.} This figure displays the monthly percentage of complaints identified as Likely-AI from a dataset of 1,134,512 CFPB complaints between March 2015 and March 2024. A vertical red dotted line indicates the release of ChatGPT on November 30, 2022. After this release, the proportion of Likely-AI complaints steadily increased, reaching a peak of 9.8\% by March 2024.
    \end{minipage}       
    \label{fig:surge_in_use}
\end{figure}

Figure \ref{fig:surge_in_use} presents a time trend of the monthly proportion of complaints that the AI detection tool identifies as likely written by an LLM. Specifically, complaints with an AI Score $\geq$ 99\% are hereafter referred to as ``Likely-AI'' complaints.\footnote{This threshold is determined based on an examination of the AI score distribution for all complaints, which exhibits a bimodal pattern: one peak occurs above 99\%, while a much more pronounced peak is observed below 5\%. More details are provided in the \textcolor{blue}{\textit{SI Appendix}, section \ref{si_subsection:ai_score_threshold_robustness}}.} In the months immediately following the initial release of ChatGPT (November 30, 2022), we observe a sharp increase in the proportion of Likely-AI complaints, from near 0\% up to 9.8\%. 

It is also notable that before the release of ChatGPT, the proportion of Likely-AI complaints was very close to zero. Although there were some instances where the AI detection tool incorrectly labeled a complaint as likely written by LLMs before ChatGPT was available (i.e., false positive), the rate was consistently negligible, near 0\%.
 
When applying less stringent AI Score thresholds (e.g., 80\% or 70\%) to define Likely-AI complaints, the rate of false positives increases. However, the surge in the share of Likely-AI complaints remains significant (see \textcolor{blue}{\textit{SI Appendix}, section \ref{si_subsection:ai_score_threshold_robustness}}). Consequently, we use a conservative threshold of 99\% AI Score in our analysis to reliably identify complaints likely generated or assisted by LLMs.

\begin{figure}[!ht]
    \centering
    \caption{LLM Adoption Patterns Varying Across Regions Based on English proficiency}
     \includegraphics[width=0.8\linewidth]{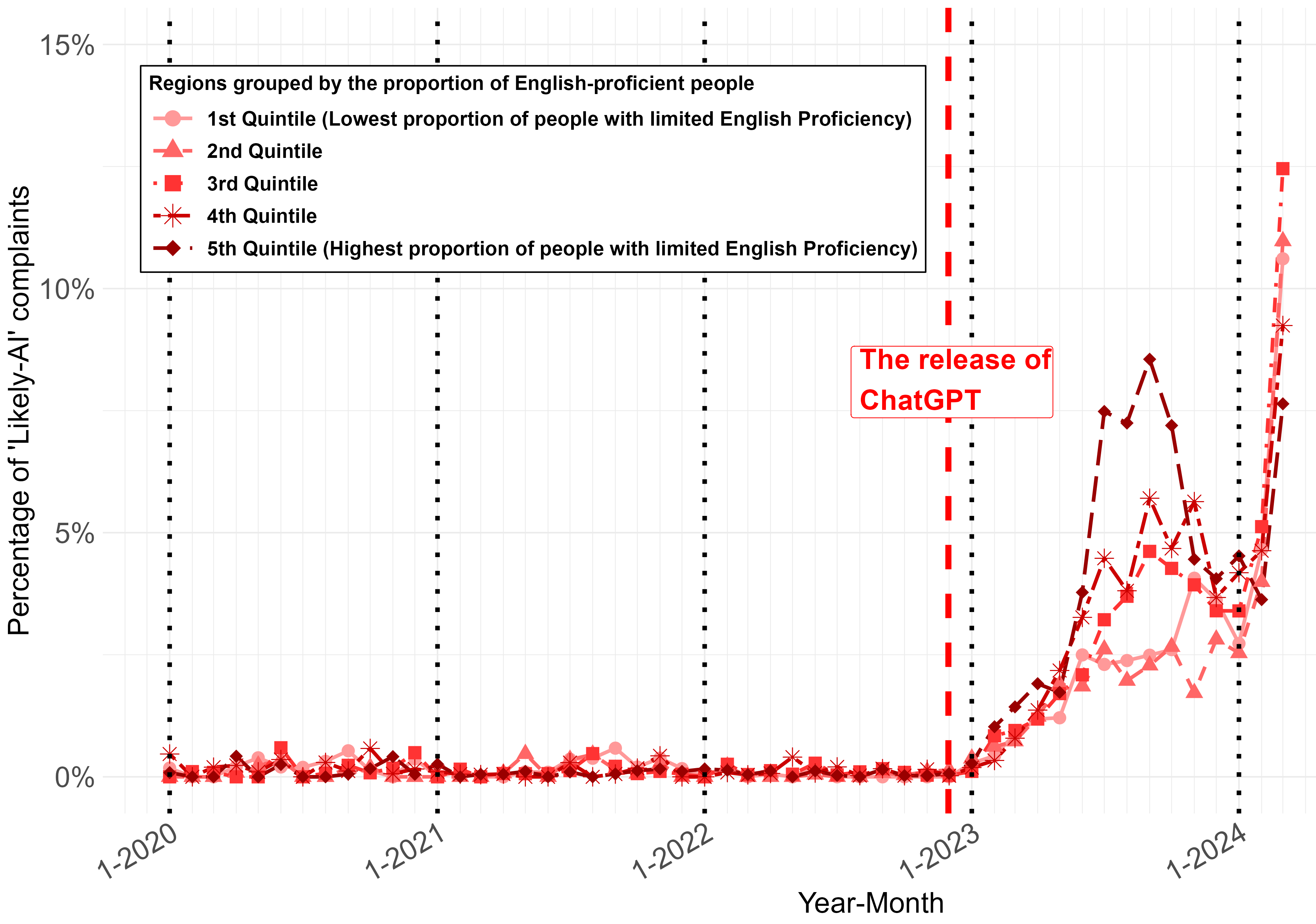}
     \begin{minipage}{0.98\linewidth}
    \vspace{0.5em}
    \small \textit{Note.} This figure displays regional variations in LLM adoption from 2020 through 2024, grouped by five English‐proficiency quintiles (based on 2017–2021 ACS data) at the ZIP‐code level. Each color–marker combination corresponds to a different quintile of ZIP codes (the deeper the red, the higher the proportion of individuals with limited English proficiency). For instance, the light pink line with circle markers (1st quintile) represents areas with the lowest average share of limited‐English‐proficient residents (around 4\%), while the bold red line with diamond markers (5th quintile) indicates the highest share (around 60\%). The figure demonstrates that early adoption of LLMs was more rapid in regions with a higher proportion of people with limited English proficiency.
     \end{minipage}     
    \label{fig:hetero_adoption}
\end{figure}

Furthermore, we observe notable regional heterogeneity in adoption patterns linked to English proficiency. For this analysis, we divided regions into five quintiles (Q1 to Q5) based on the percentage of people with limited English proficiency, using ACS data. Temporal trends were then examined separately for each quintile, as shown in Figure \ref{fig:surge_in_use}, except that the percentages are calculated based on each regional sample. As illustrated in Figure \ref{fig:hetero_adoption}, each line corresponds to one quintile, allowing us to compare the trends across regions with varying English proficiency levels.\footnote{This figure focuses on the period from 2020 to 2024 to highlight adoption patterns; the complete time trend starting from 2015 is available in the \textcolor{blue}{\textit{SI Appendix}, section \ref{si_section:full_time_trend}}.} Notably, adoption was initially concentrated in regions with greater language barriers (4th and 5th quintiles), becoming more uniformly distributed across all regions by 2024, indicating that factors beyond language barriers, such as increased awareness, likely spurred this expanded adoption. 
These findings suggest that consumers from diverse sociodemographic backgrounds might have adopted LLMs differently when submitting complaints to financial firms. This cross-sectional variation in adoption rates served as one of the main instrumental variables in our analysis of LLM efficacy.

\subsection{The Impact of Using LLMs on Getting Relief from Financial Firms}\label{subsec:obs_logistic_reg}

Having identified the Likely-AI complaints, we tested whether firms would respond more positively to those complaints. We categorized complaints submitted after the release of ChatGPT into two groups: \textit{Likely-AI complaints} (AI scores $\geq$ 99\%) and \textit{Likely-Human complaints} (AI scores $\leq$ 5\%). We then compared the outcome indicating whether the complaint received any form of relief~\textemdash~monetary or non-monetary~\textemdash~from firms between these groups. This comparison reveals a significant difference in the probability of receiving relief (Likely-AI: 49.3\% vs. Likely-Human: 39.9\%, \textit{p} \textless~.001). Even after controlling for over 200 complaint issues (i.e., distinct problems indicating what each complaint is about), and time trends, this positive association remains statistically significant; the results from various regressions with different control variables are provided in \textcolor{blue}{\textit{SI Appendix}, section \ref{si_section:additional_empirical}}.

While these results provide suggestive evidence of a positive association between LLM usage and the likelihood of obtaining relief, it is crucial to address potential biases that may arise from non-random selection into LLM usage. Specifically, consumers who choose to use LLMs may differ systematically from those who do not, potentially possessing unobserved characteristics that could influence firms' responses to complaints. Such characteristics, often correlated with socioeconomic status, could bias the estimated effect due to endogeneity, implying that the positive association may be less about the usage of LLMs and more about the socioeconomic status of the consumers.

\paragraph{Instrumental variables (IVs) estimation}

To address concerns regarding endogeneity and non-random selection into LLM usage, we used two demographic variables from the American Community Survey as our instrumental variables: \textit{Internet access} and \textit{English proficiency}, both measured at the ZIP code level.
% \footnote{These two instrumental variables are also recommended by the innovative AI-assisted search method for identifying valid instrumental variables, as proposed by \citet{han2024mining}.} 
These variables are based on 5-year estimates (2017-2021), preceding the widespread release of LLMs such as ChatGPT. The timing of these data is crucial for ensuring the exogeneity of the IVs, as it mitigates concerns that they could be contemporaneously correlated with firms' decisions to provide relief. 
Moreover, these IVs are likely to affect only the usage of LLM, without directly influencing relief outcomes. This exclusion restriction may hold particularly when controlling for other demographic variables, such as income and education.\footnote{We control for a range of observable demographic characteristics such as income, education, and employment status that might otherwise confound the relationship between LLM usage and relief outcomes. By conditioning on these covariates, we aim to ensure that the IVs are not simply capturing latent socioeconomic factors that could influence both LLM adoption and firm behavior.}

Specifically, our two IVs, Internet Access and English Proficiency, address different issues related to the potential endogeneity of ChatGPT usage. Internet Access captures the \textit{feasibility} of adopting ChatGPT: ZIP codes with robust broadband coverage, predetermined before ChatGPT's release, offer consumers a genuine opportunity to experiment with LLMs, whereas those lacking reliable connectivity remain effectively isolated. Because local infrastructure does not itself determine whether a complaint is granted relief, variation in Internet Access provides an exogenous source of change in LLM usage. In contrast, English Proficiency reflects the \textit{necessity} of using ChatGPT: consumers with weaker English skills would face a stronger incentive to rely on AI for clarity and persuasion, a need that is largely unrelated to other drivers of success (as demonstrated in our later falsification tests). Together, these two IVs enable us to address potential confounding factors and implement a two-stage estimation that isolates the impact of LLM usage on the likelihood of obtaining relief.

To test the impact of our IVs on LLM usage, we estimate the following first-stage regression:
\begin{equation}
\vspace{-0.2cm}
\ln\left( \frac{P(LLM_i = 1)}{P(LLM_i = 0)} \right)= \alpha_1 Internet_{j_i} + \alpha_2  English_{j_i} + \mathbf{X}_{j_i}' \boldsymbol{\gamma} + \delta_{t_i} + \delta_{s_i} + \delta_{p_i} +\delta_{k_i}.\label{eq:first_stage}
\vspace{0.2cm}
\end{equation}

For complaints submitted after the release of ChatGPT, this logistic regression models the log-odds that a given complaint $i$ was generated using an LLM (where $LLM_i$ is a dummy variable taking the value 1 if the complaint was generated by an LLM, and 0 otherwise); the model uses \textit{Internet access} and \textit{English proficiency} as the primary independent variables, measured at the ZIP code level ($j_i$) from 2017–2021. The model includes controls for sociodemographic factors, $\mathbf{X}_{j_i}$,  such as income, education, employment, and a set of fixed effects to account for heterogeneity across multiple dimensions. In particular, the year-month fixed effects $\delta_{t_i}$ control for broad temporal trends in both LLM adoption and relief decisions, encompassing major macroeconomic shocks such as the post-COVID environment and rising inflation after 2021.\footnote{We observe an increase in consumer complaint filings beginning around January~2022---well before ChatGPT's release. This trend underscores the importance of time-fixed effects to capture broader macroeconomic conditions that may influence complaint frequency, LLM adoption, and relief decisions.} Meanwhile, the state-level fixed effects $\delta_{s_i}$ capture cross-state heterogeneity, the product-category fixed effects $\delta_{p_i}$ account for different financial products (e.g., credit cards, mortgages), and the issue-type fixed effects $\delta_{k_i}$ control for variation in complaint issue types (e.g., billing disputes, fraud). These fixed effects help isolate the exogenous influence of the IVs on LLM usage by controlling for heterogeneity across multiple dimensions.

\begin{table}[t]
\vspace{0.3cm}
\footnotesize
\centering 
\caption{First-Stage Regression Results}
\begin{tabular}{@{\extracolsep{4pt}}lccc} 
\toprule % Top horizontal line
 & \multicolumn{3}{c}{\textit{Dependent variable: Likely-AI Dummy}} \\ 
\cmidrule{2-4} % Customized mid horizontal line starting from column 2 to 4
  & (1) & (2) & (3) \\ 
\midrule % Middle horizontal line
  Proportion of Households without Internet access  & $-$0.649$^{**}$ &&\hspace{0.2cm}$-$0.884$^{***}$ \\ 
  & (0.243) &  &(0.248)\\ 
  Proportion of People with less English proficiency  & & \hspace{0.4cm}0.249$^{***}$ &\hspace{0.4cm}0.297$^{***}$\\ 
  & & (0.063) &(0.064)\\ 
\midrule % Middle horizontal line

Year-Month Fixed Effect & Y&Y&Y\\
State Fixed Effect & Y&Y&Y\\
Complaint Issue Fixed Effect & Y&Y&Y\\
Finance Product Fixed Effect & Y&Y&Y\\
Demographic variables Included  & Y & Y & Y  \\ 
Observations & 244,407 & 244,407 & 244,407\\ 
Unit of Observations & \multicolumn{3}{c}{Each Complaint ($i$)} \\
Sample Period & \multicolumn{3}{c}{Dec-2022 to Mar-2024}\\ 
\bottomrule % Bottom horizontal line
\multicolumn{4}{r}{\textit{Note.} $^{*}$\textit{p} $<$ .05; $^{**}$\textit{p} $<$ .01; $^{***}$\textit{p} $<$ .001}
\end{tabular}
\label{tab:first_stage}
\end{table}

Table \ref{tab:first_stage} presents the estimates from our first-stage regressions. Column (1) shows that in ZIP codes with a higher proportion of households without Internet access, consumers are less likely to submit complaints classified as Likely-AI. 
% This finding remains robust when controlling for other factors such as Year-Month, states, complaint issue categories, finance product categories, and other sociodemographic variables such as total population, income, education, and employment. 
Column (2) suggests that in ZIP codes with greater language barriers, indicated by less English proficiency, consumers are more likely to submit Likely-AI complaints. Column (3) confirms the robustness of the effect of these IVs when both are included simultaneously. The first-stage results demonstrate that our IVs significantly explain the variation in whether consumers submit Likely-AI complaints, and these findings hold even when controlling for various fixed effects and additional covariates.

Leveraging exogenous variation from our two IVs in the first stage, we adopt the two-stage residual inclusion (2SRI) method outlined by \citet{terza2008two} and \citet{wooldridge2015control}. This approach, designed for models where both the endogenous regressor and the final outcome can be binary and estimated via a logistic link, uses the residual from the first stage as a control function. Specifically, we calculate standardized residuals ($\hat{r}_i\equiv \frac{LLM_i-\hat{P}(LLM_i=1)}{\sqrt{\hat{P}(LLM_i=1)(1-\hat{P}(LLM_i=1))}}$), following the idea that this can spread out the residual value to capture the rare treatment \cite{basu20182sls}.\footnote{The main motivation for our chosen residual specification is its established suitability in 2SRI settings where the binary treatment is sparsely observed (i.e., relatively few Likely-AI complaints) \cite{basu20182sls}. Given the challenge of determining \emph{a priori} which residual specification from a binary treatment model best captures the nonseparable error term in the outcome equation, researchers often consider alternative specifications like the \textit{raw residual}, $\hat{v_i}\equiv LLM_i-\hat{P}(LLM_i=1)$ (equivalent to the generalized residual under a logit model) or the \textit{deviance residual} \cite{terza2008two,wooldridge2015control,song2020capacity}. We assessed the robustness of our findings to these specifications and to the use of a linear probability model in lieu of logit, consistently confirming that LLM usage increases the likelihood of relief. Further details are provided in the \textcolor{blue}{\textit{SI Appendix}, section \ref{si_section:additional_empirical}}.} We then include \(\hat{r}_i\) in the second-stage logistic regression, as specified in Equation (\ref{eq:second_stage}), to correct for the endogeneity of LLM usage. To account for the estimation uncertainty from the first-stage regression, we employ a bootstrap procedure with 1,000 replications to estimate standard errors. 

\begin{equation}
\ln\left( \frac{P(Relief_i = 1)}{P(Relief_i = 0)} \right)= \beta_1 LLM_i + \beta_2  \hat{r}_i + \mathbf{X}_{j_i}' \boldsymbol{\gamma} + \delta_{t_i} + \delta_{s_i} + \delta_{p_i} + \delta_{k_i} \label{eq:second_stage}
\vspace{0.2cm}
\end{equation}

The main findings, presented in Table \ref{tab:second_stage}, show that consumers who used LLMs to compose their complaints are more likely to receive relief from financial firms, suggesting a potentially causal relationship.\footnote{To further test the robustness of our findings, we replicate the two-stage estimation with additional company fixed effects and results are robust and we observe a very similar relationship (see \textcolor{blue}{\textit{SI Appendix}, section \ref{si_section:additional_empirical}}).} Importantly, this relationship accounts for many potential confounders, including time trends, state-level cross-sectional differences, ZIP-code-level sociodemographic variables, and specific complaint issues and related products. Each column in Table \ref{tab:second_stage} presents results using different residuals derived from various IVs employed in the first stage.

Since the second-stage regression employs a logit specification, the coefficients represent changes in log-odds rather than linear effects. To interpret these coefficients on the probability scale, we compute the average marginal effect of the Likely-AI dummy variable. Specifically, for each complaint, we estimate the difference in the predicted probability of obtaining relief with and without the dummy variable, holding other covariates at their observed values. We then average these differences over all observations. This procedure yields an estimated increase of approximately 10.28 percentage points in the likelihood of obtaining relief when AI is used.\footnote{To further substantiate our findings, we compare complaint outcomes submitted both before and after the release of ChatGPT, focusing on regions where consumers have begun adopting LLMs (treatment group) versus those where they have not (control group). Our analysis reveals that in regions embracing LLMs, the proportion of complaints getting relief is 5 to 10 percentage points higher compared to regions where LLMs have not been widely adopted. This result supports the main conclusion of our study and is detailed in the  \textcolor{blue}{\textit{SI Appendix}, section \ref{si_section:additional_empirical}}.} 
% This result implies that the instrumental variables approach isolates plausibly exogenous variation in AI usage driven by English proficiency and Internet access, thereby uncovering an average effect for consumers whose choice to use AI is influenced by these instruments.
%%%%%%%%%%%%%%%%%%%%%%%%%%%%%%%%%%%%%%%%
\begin{table}[!ht]
\label{tab:second_stage}
\footnotesize
\centering 
\caption{Second-Stage Regression Results (with the First-Stage Control Function)}
\begin{tabular}{@{\extracolsep{4pt}}lccc} 
\toprule % Top horizontal line
 & \multicolumn{3}{c}{\textit{Dependent variable: Relief Dummy}} \\ 
\cmidrule{2-4} % Customized mid horizontal line starting from column 2 to 4
  & (1) & (2) & (3) \\ 
\midrule % Middle horizontal line
  Likely-AI Dummy & \hspace{0.3cm}0.469$^{***}$ &\hspace{0.3cm}0.473$^{***}$\hspace{-0.3cm}&\hspace{0.4cm}0.473$^{***}$\phantom{x}\\ 
  & (0.048) &(0.049)&(0.048)\\ 
  %4.688e-01  3.985e-02
  %4.725e-01  3.995e-02   
  %4.731e-01  3.989e-02
  
\midrule % Middle horizontal line

First-Stage IV & Internet access & English proficiency & Both\\
Year-Month Fixed Effect & Y&Y&Y\\
State Fixed Effect & Y&Y&Y\\
Complaint Issue Fixed Effect & Y&Y&Y\\
Finance Product Fixed Effect & Y&Y&Y\\
Demographic variables Included  & Y & Y & Y  \\ 

Observations & 244,407 & 244,407 & 244,407\\ 
Unit of Observations & \multicolumn{3}{c}{Each Complaint} \\
Sample Period & \multicolumn{3}{c}{Dec-2022 to Mar-2024}\\ 
\bottomrule % Bottom horizontal line
\multicolumn{4}{r}{\textit{Note.} $^{*}$\textit{p} $<$ .05; $^{**}$\textit{p} $<$ .01; $^{***}$\textit{p} $<$ .001}
\end{tabular}

\label{tab:second_stage}
\end{table}
\paragraph{Falsification Test} 
If these variables directly influence relief by another channel, then our causal interpretation would fail. To investigate this possibility, we conduct a falsification test using pre-ChatGPT data, a period when the AI channel was effectively absent but any alternative pathways from instruments to relief could still be present. Under our assumption, the instruments should show no predictive power for relief decisions before the advent of LLMs.\footnote{Such falsification tests are commonly employed in the instrumental variables literature to rigorously assess instrument validity (e.g., \citealt{Ananat2011Wrong, Nunn2011Slave, Lowes2021Legacy}). The main idea is to determine whether the instruments may influence the outcome through alternative pathways or whether a pre-existing relationship could violate the exclusion restriction.}

Table \ref{tab:falsification} presents the results of this falsification test, with each column corresponding to a specification analogous to those in Tables~\ref{tab:first_stage} and~\ref{tab:second_stage}. Columns (2) and (3) focus on English proficiency alone or both instruments together and show no significant relationship with relief outcomes in the pre-ChatGPT period. This supports the argument that English proficiency does not directly affect relief when LLM usage is unavailable. In contrast, Column (1), using only Internet access as the sole instrument, exhibits a marginally significant association with relief, suggesting potential alternate pathways influencing relief outcomes unrelated to AI usage. Consequently, although our falsification test suggests that English proficiency satisfies the exclusion restriction, the evidence for Internet access is less conclusive. Even so, when both instruments are included (Column (3)), we observe no significant relationship in the pre-ChatGPT period, indicating minimal direct impact from either instrument to relief decisions. This confirms the validity of English proficiency as an instrument while suggesting a more cautious interpretation for Internet access. Nonetheless, our principal findings from Table \ref{tab:second_stage} remain consistent whether we use both instruments or rely solely on English proficiency. By showing that neither instrument significantly predicts relief decisions prior to the availability of AI tools, the falsification exercise provides further confidence that our estimated effects in the post-ChatGPT era primarily operate through LLM usage.
% in the pre-ChatGPT window. This finding raises potential concerns about whether Internet access might influence relief outcomes through channels unrelated to AI usage.

\begin{table}[tbp]
\footnotesize
\centering
\caption{Falsification Test using pre-ChatGPT data}
\label{tab:falsification}
\begin{tabular}{@{\extracolsep{4pt}}lccc} 
\toprule % Top horizontal line
 & \multicolumn{3}{c}{\textit{Dependent variable: Relief Dummy}} \\ 
\cmidrule{2-4} % Customized mid horizontal line starting from column 2 to 4
  & (1) & (2) & (3) \\ 
\midrule % Middle horizontal line
   Proportion of Households without Internet access  & \hspace{0.1cm}0.191$^*$ &&0.169 \\ 
  & (0.093) &  &(0.094)\\ 
  Proportion of People with less English proficiency  & & 0.045 &0.037\\ 
  & & (0.026) &(0.027)\\ 
  
\midrule % Middle horizontal line

First-Stage IV & Internet access & English proficiency & \hspace{0.4cm}Both\phantom{xxx}\\
Year-Month Fixed Effect & Y&Y&Y\\
State Fixed Effect & Y&Y&Y\\
Complaint Issue Fixed Effect & Y&Y&Y\\
Finance Product Fixed Effect & Y&Y&Y\\
Demographic variables Included  & Y & Y & Y  \\ 

Observations & 624,588 & 624,588 & 624,588 \\ 
Unit of Observations & \multicolumn{3}{c}{Each Complaint} \\
Sample Period & \multicolumn{3}{c}{Jan-2017 to Oct-2022}\\ 
\bottomrule % Bottom horizontal line
\multicolumn{4}{r}{\textit{Note.} $^{*}$\textit{p} $<$ .05; $^{**}$\textit{p} $<$ .01; $^{***}$\textit{p} $<$ .001}
\end{tabular}

\end{table}

%paragraph to introduce the need for lab experiments
Despite these results, the IV estimation may not fully capture all the heterogeneity in consumer motivations. For instance, it is impossible in our observational data to compare otherwise-identical complaints drafted with and without LLMs, and we lack direct insight into additional consumer information (e.g., credit scores, lifetime value) that complaint handlers might consider. Moreover, this estimation primarily reflects the behavior of users particularly influenced by our instruments; it may not encompass all possible unobservable heterogeneity. To address these issues and more robustly establish causality, we complement our IV estimation with controlled lab experiments designed to directly test and validate our findings further.

\section{Controlled Experiments: Testing the Presentation-Enhancing Effect of LLM}\label{subsec:experiments}

By directly manipulating complaint content in a carefully structured environment, controlled experiments can more rigorously validate our findings from IV regressions and further explore a possible mechanism underlying firms’ differential responses to AI-generated complaints. To investigate this relationship in greater depth, we conducted three pilot studies and three preregistered experiments; see the project’s Open Science Framework (OSF) page\footnote{\textcolor{blue}{\href{https://osf.io/fh2cz/?view_only=0d1f07d0baf54cfdb8aa6fb30e02f9b3}{https://osf.io/fh2cz/?view\_only=0d1f07d0baf54cfdb8aa6fb30e02f9b3}}} for the preregistrations, materials, data, and analysis code for each of these studies.

\subsection{A Summary of Three Pilot Studies}\label{subsec:pilot_study_summary}
Three pilot studies were conducted to achieve distinct objectives: Pilot Study 1 tested whether the experimental design of our main studies accurately reflects decision-making patterns observed in CFPB data. Pilot Study 2 explored how participants might use LLMs to improve the presentation of their complaints, and Pilot Study 3 assessed the actual enhancement in complaint presentation by LLMs.  Further details on each pilot study are available in the \textcolor{blue}{\textit{SI Appendix}, section \ref{si_section:pilot_studies_and_stim_dev}}.

The primary aim of Pilot Study 1 was to confirm that the participants in our experiments could mimic the decision-making process of financial firms as seen in real CFPB data. This validation is crucial as it addresses whether our experimental setup can replicate real-world outcomes, thereby ensuring the relevance and reliability of our experimental observations. 

To assess this, we carefully selected a random set of 108 complaints from the CFPB data, all written before the release of LLMs (i.e., before 2022). Half of these complaints (\textit{n} = 54) had resulted in actual monetary relief from firms, while the other half (\textit{n} = 54) had not resulted in any relief from firms.
% we ensured that these complaints represented the top three financial products and issues.
We then recruited 216 participants from Prolific who each evaluated a random set of 5 complaints from these complaints. Participants rated the likelihood of offering monetary compensation for each complaint on a scale from 1 (\textit{Not likely at all}) to 7 (\textit{Extremely likely}). 
The results closely aligned with the complaints' actual outcomes: Participants were more likely to offer hypothetical compensation for the complaints that had resulted in actual monetary relief from firms (\textit{M} = 4.75, \textit{SD} = 1.84) than for the complaints that had not (\textit{M} = 4.20, \textit{SD} = 1.98), \textit{t} = 4.74, \textit{p} < .001. This result supports the validity of our experimental design and suggests that our participants' evaluation of complaints mirrors that of actual decision-makers in financial firms.

Pilot Study 2 examined how people might use an LLM to enhance the presentation of their complaints. We recruited 212 participants from Prolific and asked them to imagine writing complaints ``to obtain monetary relief from a financial institution.'' Each participant reviewed five randomly selected complaints from 150 CFPB complaints, rating their intention to improve 10 linguistic features, such as politeness and assertiveness. The results showed a strong preference for enhancing clarity, coherence, and professionalism (see \textcolor{blue}{\textit{SI Appendix}, section \ref{si_subsection:pilot_study_2}} and \textcolor{blue}{Table \ref{tab:top_3_ling_features}}). Using these findings,
 
we developed stimuli for the main experiments. We selected 20 different complaints from the CFPB data and employed ChatGPT to refine each for greater clarity, coherence, or professionalism, ensuring that improvements were strictly stylistic, and kept the complaints’ content identical across the edited and unedited versions (see \textcolor{blue}{\textit{SI Appendix}, section \ref{si_subsection:stimuli_development}}). This resulted in 80 complaints, comprising 20 unedited (\textit{Control}) and 60 edited (20 \textit{Clear}, 20 \textit{Coherent}, and 20 \textit{Professional}). These complaints were then used as stimuli in the main experiments.

Pilot Study 3 tested whether these 60 edited complaints were indeed enhanced in terms of overall presentation style compared to the 20 unedited ones. We recruited 490 participants from Prolific and randomly assigned them to one of three conditions: Clarity, Coherence, or Professionalism. Each participant received a series of five complaints randomly selected from the pool of 80 complaints (20 Control, 20 Clear, 20 Coherent, and 20 Professional), ensuring no overlap in content. They rated each complaint on the linguistic feature of their assigned condition.

% \vspace{0.5cm}

\begin{figure}
    \centering
    % \caption{Editing Complaints With an LLM Enhanced the Presentation}
    \caption{Enhancement in Presentation for the LLM-Edited vs. Unedited Complaints}
    \includegraphics[width=\textwidth]{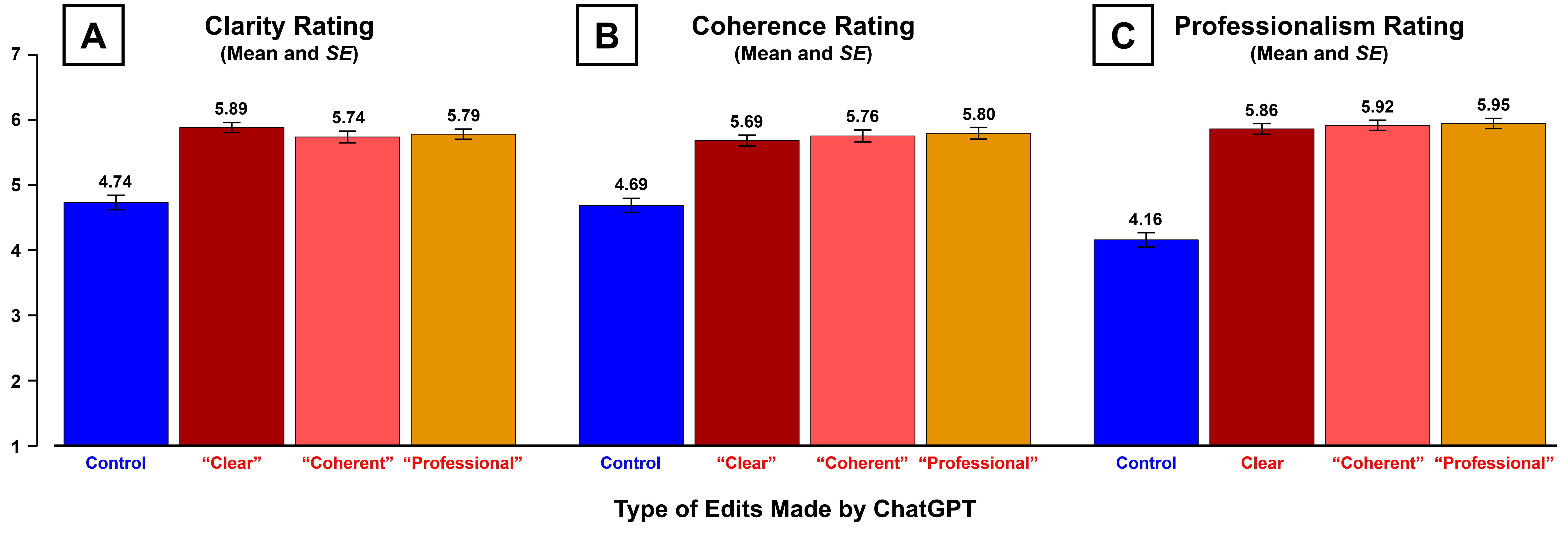}
    \begin{minipage}{0.98\linewidth}
    \vspace{1em}
    \small \textit{Note.} Panels A-C display results from Pilot Study 3. Edited complaints (Clear, Coherent, and Professional complaints) showed enhanced presentation compared to the unedited complaints (Control). 
    % The enhancement in presentation was statistically significant in pairwise \textit{t}-tests as well as in the preregistered linear fixed effects models that controlled for the fixed effects of individual participants and complaint contents.
    
    \end{minipage}    
    \label{fig:pilot_study_results}
    \vspace{0.1cm}
\end{figure}

Results revealed that the 60 edited complaints demonstrated enhancements in \textit{all three} linguistic features despite each participant assessing only one feature, suggesting that LLM editing improves the overall presentation quality of complaints across multiple linguistic features simultaneously \citep{niven2019probing, manning2020emergent}. The validation of these presentation enhancements in LLM-edited complaints confirms their validity for use in our main experiments, facilitating an analysis of how stylistic improvements affect decisions on financial relief.

\subsection{Experiment 1}\label{subsec:exp_1}
In Experiment 1 (\textit{N} = 301), we presented each participant with five distinct complaints, randomly selected from the set of 80, 
% ensuring that each set of complaints appeared only once to avoid content repetition 
ensuring that the contents of complaints were never repeated for a given participant
(for more details, see \textcolor{blue}{\textit{SI Appendix}, section \ref{si_subsection:exp_1}}). Participants rated the likelihood of providing monetary compensation in response to each complaint on a scale from 1 (\textit{Not likely at all}) to 7 (\textit{Extremely likely}), and this response served as the main dependent measure of the experiment. 

Participants then provided feedback on the survey (optional) and demographic information to complete the study.

Results of Experiment 1 align with our findings from the CFPB data. We primarily focus on the preregistered linear fixed effects model  (Model 2 in Table \ref{tab:lfe_models_in_exp_1}), which controls for the effects of participants and complaint contents. This model showed that the complaints edited with ChatGPT (\textit{Coherent}, \textit{Clear}, and \textit{Professional} complaints) were more likely to receive hypothetical monetary compensation than the unedited (\textit{Control}) complaints, \textit{b} = 0.31, \textit{SE} = 0.087, \textit{t} = 3.52, \textit{p} < .001; see Model 2 in Table \ref{tab:lfe_models_in_exp_1}.
Additional linear models (Models 1, 3, and 4), controlling for different sets of fixed effects, consistently indicate that editing with ChatGPT significantly increased the likelihood of hypothetical compensation, regardless of the specific linguistic feature edited. These results corroborate the findings from Section \ref{subsec:obs_logistic_reg}, confirming that editing complaints with an LLM significantly increases the likelihood of obtaining relief.

We also conducted a nearly direct replication of this experiment, which showed the same pattern of results, all with greater statistical significance (all \textit{p}s < .001). This preregistered experiment is reported as ``Experiment S1'' in the \textcolor{blue}{\textit{SI Appendix}, section \ref{si_subsection:exp_s1}}.

\vspace{0.3cm}

% \begin{table}[!ht]
\begin{table}[htbp]
\centering 
\footnotesize
\caption{Results From Experiment 1}
\label{tab:lfe_models_in_exp_1}

\begin{tabular}{@{\extracolsep{5pt}}l 
                >{\raggedright\arraybackslash}m{1.3cm} 
                >{\raggedright\arraybackslash}m{1.3cm} 
                >{\raggedright\arraybackslash}m{1.3cm} 
                >{\raggedright\arraybackslash}m{1.3cm}} 
\\[-1.8ex]\hline 
\hline \\[-1.8ex] 

& \multicolumn{4}{c}{\textit{Dependent variable: Compensation likelihood}} \\ 
\cline{2-5} 
\\[-1.8ex] 
& \multicolumn{4}{c}{Model} \\ 
\\[-1.8ex] & \multicolumn{1}{c}{(1)} & \multicolumn{1}{c}{(2)} & \multicolumn{1}{c}{(3)} & \multicolumn{1}{c}{(4)} \\ 
\hline \\[-1.8ex] 

Constant 
& 3.645$^{***}$ & & & \\ 
& (0.104)       & & & \\ 
Unedited vs. Edited (Dummy variable) 
& 0.274$^{*}$ & 0.306$^{***}$ & 0.308$^{***}$ & \\ 
& (0.121)     & (0.087)       & (0.087)      & \\ 

Unedited vs. More clear (Dummy variable)
& & & & 0.261$^{*}$\\ 
& & & & (0.107) \\ 

Unedited vs. More coherent (Dummy variable) 
& & & & 0.279$^{**}$\\ 
& & & & (0.106) \\ 

Unedited vs. More professional (Dummy variable) 
& & & & 0.383$^{***}$\\ 
& & & & (0.107) \\ 

\hline \\[-1.8ex] 
Complaint Content (20 Different Contents) Fixed Effects 
& \multicolumn{1}{c}{N} & \multicolumn{1}{c}{Y} & \multicolumn{1}{c}{Y} & \multicolumn{1}{c}{Y}\\
Participant Fixed Effects 
& \multicolumn{1}{c}{N} & \multicolumn{1}{c}{Y} & \multicolumn{1}{c}{Y} & \multicolumn{1}{c}{Y}\\
Complaint Presentation Position (1--5) Fixed Effects 
& \multicolumn{1}{c}{N} & \multicolumn{1}{c}{N} & \multicolumn{1}{c}{Y} & \multicolumn{1}{c}{Y}\\
Observations 
& \multicolumn{1}{c}{1,505} & \multicolumn{1}{c}{1,505} & \multicolumn{1}{c}{1,505} & \multicolumn{1}{c}{1,505}\\
\hline 
\hline \\[-1.8ex] 

\multicolumn{5}{p{\dimexpr\linewidth-2\tabcolsep}}{\raggedright \textit{Note.} $^{*}$\textit{p} $<$ .05; $^{**}$\textit{p} $<$ .01; $^{***}$\textit{p} $<$ .001. We preregistered estimation of Model 2 as the analysis to test our main hypothesis, as can be seen in the preregistration available on the \textcolor{blue}{\href{https://osf.io/fh2cz/?view_only=0d1f07d0baf54cfdb8aa6fb30e02f9b3}{OSF page}}.}

\end{tabular} 
\end{table}

% \paragraph{Robustness and Relevance Tests}
\par\bigskip % Adds a natural space without forcing a new page
\subsection{Experiment 2}\label{subsec:exp_2_methods_and_results}

Experiment 2 (\textit{N} = 400) was conducted with participants who had \textit{prior work experience in the finance industry}, thereby enhancing the external validity of our findings. Participants were asked to evaluate complaints based on a specific presentation feature, either clarity or professionalism (see \textcolor{blue}{\textit{SI Appendix}, \ref{si_subsection:exp_2}}). The results replicated the findings from the previous experiments, confirming that the observed effects were robust in a more ecologically valid sample and were consistent even when participants focused on a specific presentation feature (clarity or professionalism). Hence, the results from all experiments demonstrate that LLMs can improve the effectiveness of complaints by enhancing their presentation.
% , thereby catering to the diverse preferences of complaint handlers across different linguistic dimensions. 

\section{Conclusion and Implications}

%summary 
Large language models are among the most transformative advancements in AI technology. Despite their widespread influence, the degree to which consumers adopt LLMs for interacting with firms, as well as the actual efficacy of LLMs in consumer-firm communication, both remain significant empirical questions. This paper contributes as one of the first empirical studies to explore and document the rapid adoption of LLMs by consumers for drafting complaints to financial firms following the release of ChatGPT. Our analysis of over 1 million complaints from the CFPB database reveals a sharp and significant increase in LLM usage, which is also positively associated with a higher probability of obtaining relief. To address concerns about endogeneity and non-random selection into LLM usage, we employ instrumental variables (specifically, Internet access and English proficiency) that are presumed to influence LLM usage without directly affecting relief outcomes. This analysis provides plausibly causal evidence that Likely-AI complaints have an increased likelihood of securing financial relief. To substantiate this causal interpretation, we complement the CFPB data analysis with experimental validation. These lab experiments demonstrate that improvements in textual presentation by LLMs significantly enhance the clarity and persuasiveness of consumer communications, thereby increasing the likelihood of obtaining financial relief.

The findings from this study carry significant implications for consumer advocacy, firm practices, and regulatory policies. By facilitating access to LLM technologies such as integrating LLM-aided tools on their websites where customers submit complaints, firms can enhance consumer satisfaction, particularly aiding those who may struggle with articulating their complaints effectively. Leveraging LLMs to improve communication allows firms to focus on the actual content of disputes rather than being influenced by presentation style. This approach has the potential to identify previously overlooked consumer segments whose concerns were less vocal or poorly articulated, leading to more effective customer service and complaint resolution strategies. Ultimately, this fosters stronger relationships between firms and consumers, ensuring that genuine issues are addressed more efficiently.

From a policy perspective, equitable access to LLM technology is crucial to prevent exacerbating existing disparities in consumer outcomes. As our research suggests, LLM usage can significantly enhance the likelihood of obtaining relief from firms, primarily due to the improved presentation style and clarity provided by these tools. However, without equal access, consumers facing language barriers or those with less tech-savvy may find themselves at a disadvantage. Policymakers should consider strategies to broaden access to LLM tools, such as making them available through public libraries, consumer advocacy groups, or other community resources.

Moreover, regulatory bodies like the CFPB should play a proactive role in ensuring that LLM technologies are accessible to all consumers, especially those from vulnerable groups. By promoting the integration of these tools across socioeconomic segments, regulators can help level the playing field and ensure that all consumers have equal opportunities to effectively voice their concerns, enhancing the overall fairness and effectiveness of financial dispute resolutions.

\clearpage

%
% \end{center}
\bibliographystyle{chicago}
\bibliography{reference}  

\clearpage

\appendix
\newpage
\setcounter{page}{1}
\begin{center}
\textbf{\Large SI Appendix for \\The Adoption and Efficacy of Large Language Models: Evidence From Consumer Complaints in the Financial Industry}
\end{center}

\renewcommand{\thesection}{\arabic{section}}
\renewcommand{\thesubsection}{\thesection\Alph{subsection}}
\setcounter{section}{0}

\section{Empirical Study using CFPB data}
\label{si_section:obs_studies_overview}

% new commands for numbering tables and figures
\renewcommand{\thetable}{S\arabic{table}}
\setcounter{table}{0} % Reset the table counter
\renewcommand{\thefigure}{S\arabic{figure}}
\setcounter{figure}{0} % Reset the figure counter
In this paper, we examine both the adoption and efficacy of large language models (LLMs) by consumers in drafting complaints to financial firms. Specifically, we address the following research questions: (1) Are consumers adopting LLMs to draft complaints that can lead to obtaining relief from financial firms (Adoption)? (2) Does the usage of LLMs increase the likelihood of consumers receiving relief (Efficacy)? (3) Is there an interesting and novel mechanism behind this increased likelihood?

Investigating these questions presents two significant challenges. First, there is no existing data that explicitly indicates whether consumers are using LLMs to write their complaints. Large-scale datasets do not record LLM usage, and complaints typically lack any markers indicating the involvement of an LLM. Second, few large-scale, historical datasets provide both the full text of consumer complaints and detailed information about the corresponding outcomes from companies. Moreover, to observe changes in efficacy after the release of ChatGPT, we require a dataset that spans a substantial period before and after its introduction.

To overcome these challenges, we employed a recently developed LLM usage detection tool and utilized a unique, publicly available dataset. Specifically, for the first challenge, we used Winston AI, a state-of-the-art AI detection program designed to estimate the probability that a given text was generated by an LLM such as ChatGPT. This tool allowed us to identify which complaints were likely written with the assistance of an LLM versus those composed solely by humans. Further discussion about this detection tool is presented in Section \ref{si_subsection:ai_detection_tool}.

For the second challenge, we analyzed data from the Consumer Financial Protection Bureau (CFPB) Consumer Complaint Database, supplemented with sociodemographic information from the American Community Survey (ACS). The CFPB dataset provides a comprehensive historical record of over 1 million consumer complaints submitted between 2015 and 2024, along with detailed outcomes regarding the firms' responses. This rich dataset enables us to systematically examine both the content of the complaints and the corresponding outcomes over a significant time span, encompassing periods before and after the release of ChatGPT. By leveraging this dataset, we can assess changes in the efficacy of complaints potentially associated with LLM usage.

The remainder of the first section of this appendix, which supplements our observational analyses, is organized as follows. In Sections \ref{si_subsection:data_cfpb} and \ref{si_subsection:data_acs}, we introduce the two datasets used in our analysis and provide detailed descriptions of their contents and characteristics. In Section \ref{si_subsection:ai_detection_tool}, we explain the AI detection tool, discuss its performance, and cover other related aspects of our study.

% \subsection*{Data: CFPB complaint database}
% \subsection*{1A. Data: CFPB Complaint Database}
\subsection{Data: CFPB Complaint Database}
\label{si_subsection:data_cfpb}

\paragraph{CFPB Complaint Process Overview \& Data} 
The CFPB allows consumers to submit complaints regarding various financial products and services, such as credit cards, mortgages, student loans, and checking accounts. The CFPB website provides a platform for consumers to voice their concerns, and the CFPB facilitates communication between the consumer and the company involved. First, consumers submit complaints through the CFPB website or other channels (e.g., phone, mail). The consumer must provide key details, including the company involved, the issue, and any supporting documentation. The CFPB forwards the complaint to the company, allowing it to respond. If another government agency is better suited to handle the issue, the CFPB forwards the complaint to them and informs the consumer. The company typically responds within 15 days. If more time is needed, the company may provide an interim response and finalize the resolution within 60 days. Once the company responds, the consumer is notified and can review the response. The consumer also has the option to dispute the response if they are unsatisfied. Complaints and company responses are published in the CFPB database, with personal information removed to protect consumer privacy.

The CFPB database contains the following information as part of this complaint process:

\begin{itemize}[noitemsep]
    \item \textbf{Complaint ID}: A unique identifier for each complaint.
     \item \textbf{Consumer complaint narrative}: Detailed description of the complaint in the consumer’s words.

    \item \textbf{Product}: The type of financial product involved in the complaint (e.g., mortgage, credit card). There are 21 unique products in our dataset.
    \item \textbf{Sub-product}: A more specific category under the main product (e.g., FHA mortgage, checking account). There are 86 unique sub-products in our dataset.
     \item \textbf{Issue}: The specific issue raised by the consumer (e.g., billing disputes, loan modification). There are 172 unique issues in our dataset.
     \item \textbf{Sub-issue}: Further detail about the issue (e.g., incorrect fees or charges). There are 266 unique sub-issues in our dataset.
    \item \textbf{Submitted via}: The channel through which the complaint was submitted (e.g., web, phone, postal mail).
    \item \textbf{Date received}: The date the complaint was received by the CFPB.
    \item \textbf{State}: U.S. state where the consumer resides.
    \item \textbf{ZIP code}: The consumer’s ZIP code. There are 6,932 unique five-digit ZIP codes in our dataset. 
    \item \textbf{Consumer consent provided}: Whether the consumer consented to publish their narrative.
    \item \textbf{Date sent to company}: The date the CFPB forwarded the complaint to the company.
    \item \textbf{Company}: The company involved in the complaint. There are 5,625 unique companies in our dataset. 
    \item \textbf{Company response}: How the company responded to the complaint (e.g., closed with monetary relief, closed with explanation).
    \item \textbf{Timely response?}: Whether the company responded within the required timeframe.\footnote{We do not find any meaningful change in this variable before and after the release of ChatGPT; it fell below 1\% before the release and remained fairly stable around 1\%.} 
    \item \textbf{Consumer disputed?}: Whether the consumer disputed the company’s response.\footnote{This would be very useful for further measuring changes in consumer satisfaction when using LLMs; however, our sample, selected based on the observability of complaint texts, shows no variation in this variable after April 2017.}
\end{itemize}

Specifically, the firm chooses a category that best describes its response. The firm can choose from several options to categorize its response to the complaint. One option is ``Closed with monetary relief,'' which implies that the firm's response included providing objective, measurable, and verifiable monetary relief to the consumer. Another category is ``Closed with non-monetary relief,'' which indicates that the firm's actions did not result in monetary relief, but may have addressed some or all of the consumer’s complaints beyond mere explanations (e.g., changing account terms or coming up with a foreclosure alternative). A third category is ``Closed with Explanation,'' which indicates that the firm simply explained the issue without providing either monetary or non-monetary relief. In the present research, we define the successful persuasion outcome to be receiving either monetary or non-monetary relief, consistent with the definitions used in previous research \cite{dou2023learning}. This definition is used to code for the main dependent variable in our analyses. 

%\paragraph{Additional data patterns from CFPB dataset}

%Data selection (text is not available for every complaint)
The original CFPB complaint database, which is publicly accessible, contains complaints from December 2011 through April 2024 and continues to receive more complaints. However, it is important to note that not every consumer has disclosed the ``narrative,'' which is the primary text-based information of the complaint. If certain consumers do not provide consent to make their complaints public, we are unable to view the text within those complaints. This results in our data starting from March 2015 (i.e., we do not have access to complaint texts before March 2015). In addition, a significant number of complaints from April 2024 had not yet completed the complaint handling process, so we did not observe the company responses. As a result, our data ends in March 2024. Finally, the AI detection tool we utilized has specific requirements concerning the length of the input text. It requires a minimum of 500 characters to analyze, so we removed complaints that were too short to determine whether they were written using LLMs. We finalized our dataset by considering these factors, which left us with more than 1.1 million complaints ($N$ = 1,134,512) from March 2015 to March 2024.

\subsection{Data: American Community Survey Database}
\label{si_subsection:data_acs}
We construct two instrumental variables, \textit{Internet access} and \textit{English proficiency}, using data from the American Community Survey 5-Year Estimates for 2021. The data were retrieved from the U.S. Census Bureau's online platform (\url{https://data.census.gov}), focusing on subject tables pertinent to each variable.

\paragraph{Internet access} To construct the first instrumental variable, \textit{Internet access}, we utilized Table S2801 named \emph{``Types of Computers and Internet Subscriptions.''} From this table, we extracted the following variables at the five-digit ZIP code level:

\begin{itemize}[leftmargin=2em]
\item \textbf{Total Households (\texttt{S2801\_C01\_001E}):} This variable represents the estimate for the total number of households in each ZIP code (\emph{``Estimate!!Total!!Total households''}). 
\item \textbf{Households Without Internet Subscription (\texttt{S2801\_C01\_019E}):} This variable indicates the number of households without any Internet subscription in each ZIP code (\emph{``Estimate!!Total!!Total households!!TYPE OF INTERNET SUBSCRIPTIONS!!Without an Internet subscription''}).
\end{itemize}

The proportion of households with limited Internet access in each ZIP code was calculated using the formula:

\begin{equation}
\text{IV}_{(\text{No Internet access})} = \frac{\text{The number of Households without Internet}}{\text{Total Households}}
\end{equation}

This ratio represents the share of households lacking Internet access, serving as our first instrument.

\paragraph{English proficiency Instrumental Variable} For the second instrumental variable, \textit{English proficiency}, we used Table S1601 named \emph{``Language Spoken at Home.''} The variables extracted at the ZIP code level include:

\begin{itemize}[leftmargin=2em]
\item \textbf{Total Population Aged 5 Years and Over (\texttt{S1601\_C01\_001E}):} This variable provides the estimate for the total population aged 5 years and over in each ZIP code (\emph{``Estimate!!Total!!Population 5 years and over''}).     
\item \textbf{Population Speaking a Language Other Than English (\texttt{S1601\_C01\_003E}):} This variable denotes the number of people who speak a language other than English at home (\emph{``Estimate!!Total!!Population 5 years and over!!Speak a language other than English''}).
\end{itemize}

In each ZIP code, we defined the proportion of people with potentially limited English proficiency as follows:

\begin{equation}
\text{IV}_{(\text{Limited English proficiency})}= \frac{\text{Population Speaking Other Languages}}{\text{Total Population}}
\end{equation}
This proportion serves as our second instrument to capture variations in English language proficiency across ZIP codes.

In addition to these instrumental variables, we control for other demographic factors:
\begin{itemize}[leftmargin=2em]
    \item \textbf{Median Household Income:} (\emph{``Estimate!!Households!!Median income''}; \texttt{S1901\_C01\_012E})
    \item \textbf{Educational Attainment:} (\emph{``Estimate!!Percent!!AGE BY EDUCATIONAL ATTAINMENT!!Population 25 years and over!!Bachelor's degree or higher''}; \texttt{S1501\_C02\_015E})
    \item \textbf{Employment Rate:} (\emph{``Percent!!EMPLOYMENT STATUS!!Population 16 years and over!!In labor force!!Civilian labor force!!Employed''}; \texttt{DP03\_0004PE})
\end{itemize}

% \subsection*{Method: AI Detection Tool}
\subsection{Method: AI Detection Tool}
\label{si_subsection:ai_detection_tool}

In our research, we utilized Winston AI, one of the most effective AI detection programs, to determine whether a complaint was written with an LLM. This tool's performance has been tested and verified by many users and researchers \cite{prillaman2023detector,weber2023testing}, who have deemed it one of the most effective AI detection tools, on par with \textit{Originality.ai} \cite{originalityai}. Notably, this program is designed and trained to detect outputs by a wide range of popular LLMs including GPT-3, GPT-4, Gemini, Bard, Claude, and many others.

Although the developers of Winston AI have not disclosed the exact algorithm used to calculate the score, they have provided insight into the high-level principles behind it.\footnote{This explanation of how Winston AI calculates the score is from Winston AI, available at: \url{https://gowinston.ai/interpreting-our-ai-detection-scores/}} Winston AI relies on a large-scale proprietary dataset containing both AI-generated and human-written texts, which allows the system to know the ``ground truth'' for each sample. Using this dataset, Winston AI trains its model to recognize patterns that distinguish AI-authored content from human-authored content. The detection process is based on two primary methods: linguistic analysis and comparison with previously known AI-generated texts. Through linguistic analysis, the system examines specific characteristics, such as semantic structure and repetition, to detect signs of AI authorship. Two key metrics in this analysis are \textit{Perplexity} and \textit{Burstiness}. Perplexity measures how well a language model predicts the next word in a sequence, with AI-generated texts generally exhibiting lower perplexity due to the model encountering similar patterns during training. In contrast, more complex and unpredictable texts are likely human-written. Burstiness evaluates the repetition of words and phrases within a text; AI models tend to overuse certain terms, resulting in clusters of repetition, whereas human writers typically show more variation. By analyzing these linguistic features and comparing the text against known AI-generated samples, Winston AI can reliably assess whether a piece of content is AI-generated or human-written.

To verify the accuracy of Winston AI, we conducted our own tests using the CFPB database. We have a solid understanding that prior to December 2022, very few, if any, complaints would have been written with the assistance of LLMs currently available such as ChatGPT\footnote{Before the end of 2022 when ChatGPT was released, LLMs from the other GPT-series (e.g., text-ada-001 or text-curie-001) were released but not widely used. This limited usage was primarily due to the limited access and a general lack of public understanding about them.}. Thus, we can confidently categorize these complaints as human-authored, serving as ground truth for our test. We randomly selected 100 complaints from this group. We then created counterfactual, LLM-edited versions of these complaints using ChatGPT and categorized them as LLM-authored. The two groups of complaints (human-authored and LLM-authored) were processed through Winston AI as a test of the the tool's detection accuracy. The results effectively illustrate the tool's capability to differentiate between human-authored and AI-authored complaints. For human-authored complaints, the AI scores are significantly lower (first quartile: 0.010, median: 0.380, mean: 5.097, third quartile: 2.042). On the other hand, AI-authored complaints have much higher scores (first quartile: 95.65, median: 99.69, mean: 93.42, third quartile: 99.94). The stark difference in these sets of AI scores underscores the tool's effectiveness in distinguishing between complaints authored by humans and those authored by LLMs.

To further assess the robustness of our detection results, we employ another leading AI detection program, \textit{Originality.AI}. Due to the substantially higher expense associated with this program, we limit our analysis to the one-year period surrounding the release of ChatGPT (April 2022 to March 2023). First, we observe a high degree of agreement between the two AI detection programs in their binary classification labels, with an agreement rate of 94.1\% when using a 1\% cutoff in both programs to define ``Likely-AI'' complaints. Second, \textit{Originality.AI} exhibits a pattern of LLM adoption similar to our earlier observations. Specifically, prior to December 2022, the proportion of Likely-AI complaints averaged around 3\%, showing only minor fluctuations. However, beginning in December 2022, this proportion increased more noticeably, reaching 7\% shortly thereafter. This upward trend continued steadily, reaching 11\% by March 2023.

\subsection{Robustness Test of Various Thresholds for Determining Likely-AI Complaints}
\label{si_subsection:ai_score_threshold_robustness}
\begin{figure}[!htp]% [hpbt] what you need
    \centering
     \includegraphics[width=0.6\linewidth]{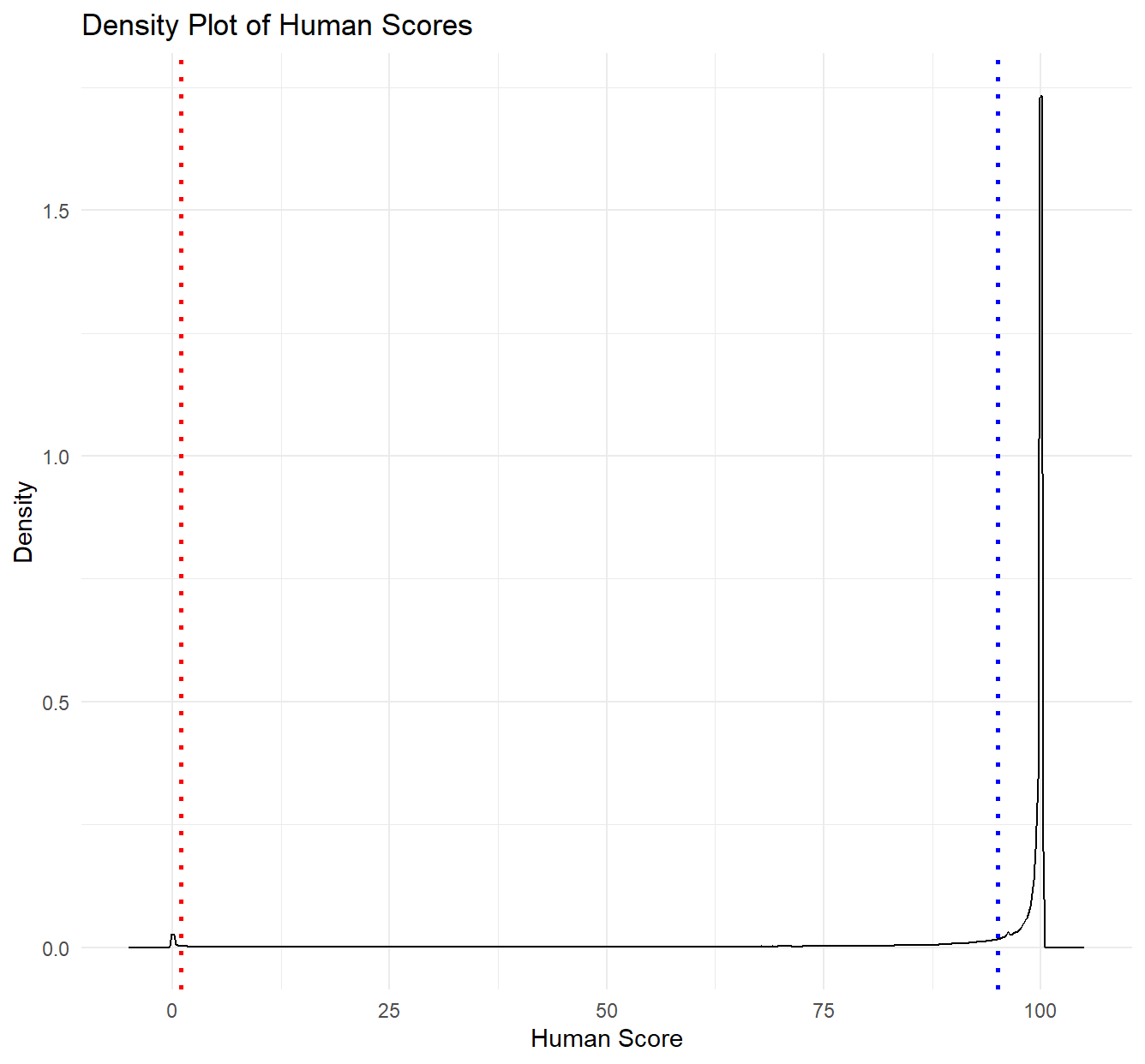}
    
     \caption{This figure presents the density distribution of human scores, defined as 100 - AI score, across the entire dataset. It reveals a bimodal distribution, with two distinct peaks: one concentrated near 0 and the other near 100. These peaks suggest two predominant classifications within the dataset. The blue vertical dotted line marks a human score threshold of 95, while the red vertical dotted line indicates a threshold of 1. Based on this distribution, we classify complaints as ``Likely-Human'' when the human score exceeds 95\% (i.e., AI score $\leq$ 5\%), and ``Likely-AI'' when the human score falls below 1\% (i.e., AI score $\geq$ 99\%). The density plot thus provides a visual foundation for our classification thresholds, highlighting the natural separation between two groups of complaints. }
    
    \label{fig:ai_score_density}
\end{figure}

We present here the rationale behind selecting 95\% as the threshold for classifying Likely-AI complaints. First, Figure \ref{fig:ai_score_density} illustrates the density distribution of the output generated by the AI detection program we employed (Winston AI), which estimates the likelihood that the input text was produced by a human rather than LLMs. The program assigns a human score to each text. The figure reveals a bimodal distribution, with two distinct peaks: one centered near 0 and the other near 100. These peaks suggest the presence of two predominant classifications within the dataset. The blue vertical dotted line marks a human score threshold of 95, while the red vertical dotted line represents a threshold of 1. Based on this distribution, we classify complaints as ``Likely-Human'' when the human score exceeds 95\% (i.e., AI score $\leq$ 5\%), and ``Likely-AI'' when the human score falls below 1\% (i.e., AI score $\geq$ 99\%). This visualization provides the foundation for our classification thresholds, highlighting the natural division between two distinct groups of complaints.

\begin{figure}[!htp]% [hpbt] what you need
    \centering
     \includegraphics[width=0.65\linewidth]{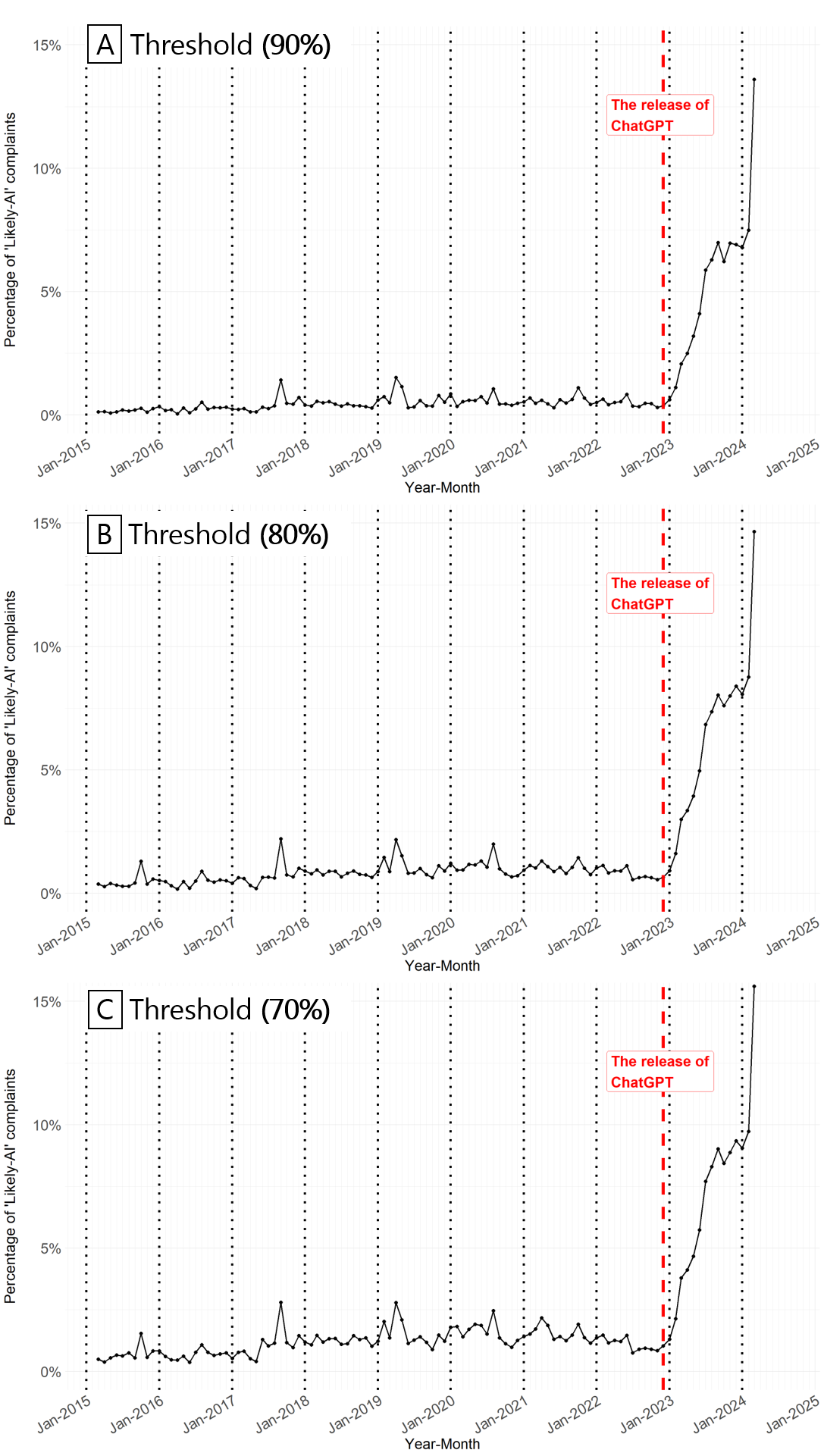}
     \caption{Each panel in the series displays the temporal trend for ``Likely-AI'' complaints, defined using various threshold levels. The top panel represents cases classified as Likely-AI when the AI score is greater than or equal to 90\%. The middle panel corresponds to cases with an AI score threshold of greater than or equal to 80\%, and the bottom panel includes cases with AI scores greater than or equal to 70\%. More false positive cases are observed with a lower threshold; that is, when the threshold is less stringent, the percentage of ``Likely-AI'' complaints is more likely to be greater than zero before the release of ChatGPT. }
     %\hspace{1.0cm}
     %\subfloat[Comparative analysis of linguistic features in Likely-AI complaints versus human complaints, illustrating significant differences in coherence, politeness, sentiment, and readability.]{\includegraphics[width=0.41\linewidth]{Figures/example.png}}
    %\caption{Likely-AI Complaints: Trends and Linguistic Analysis} 
    \label{fig:ai_score_threshold_robustness}
\end{figure}

Moreover, Figure \ref{fig:ai_score_threshold_robustness} presents another reason to use the selection criterion of an AI score $\geq$ 99\%. It clearly shows that there is a sharp increase in Likely-AI complaints immediately following the release of ChatGPT, regardless of the threshold applied. However, a less stringent threshold leads to a higher incidence of false positives when classifying Likely-AI complaints.

\newpage
\subsection{Heterogeneity in LLM Adoption (2015-2024)}
\label{si_section:full_time_trend}

\begin{figure}[!ht]
    \centering

     \includegraphics[width=1\linewidth]{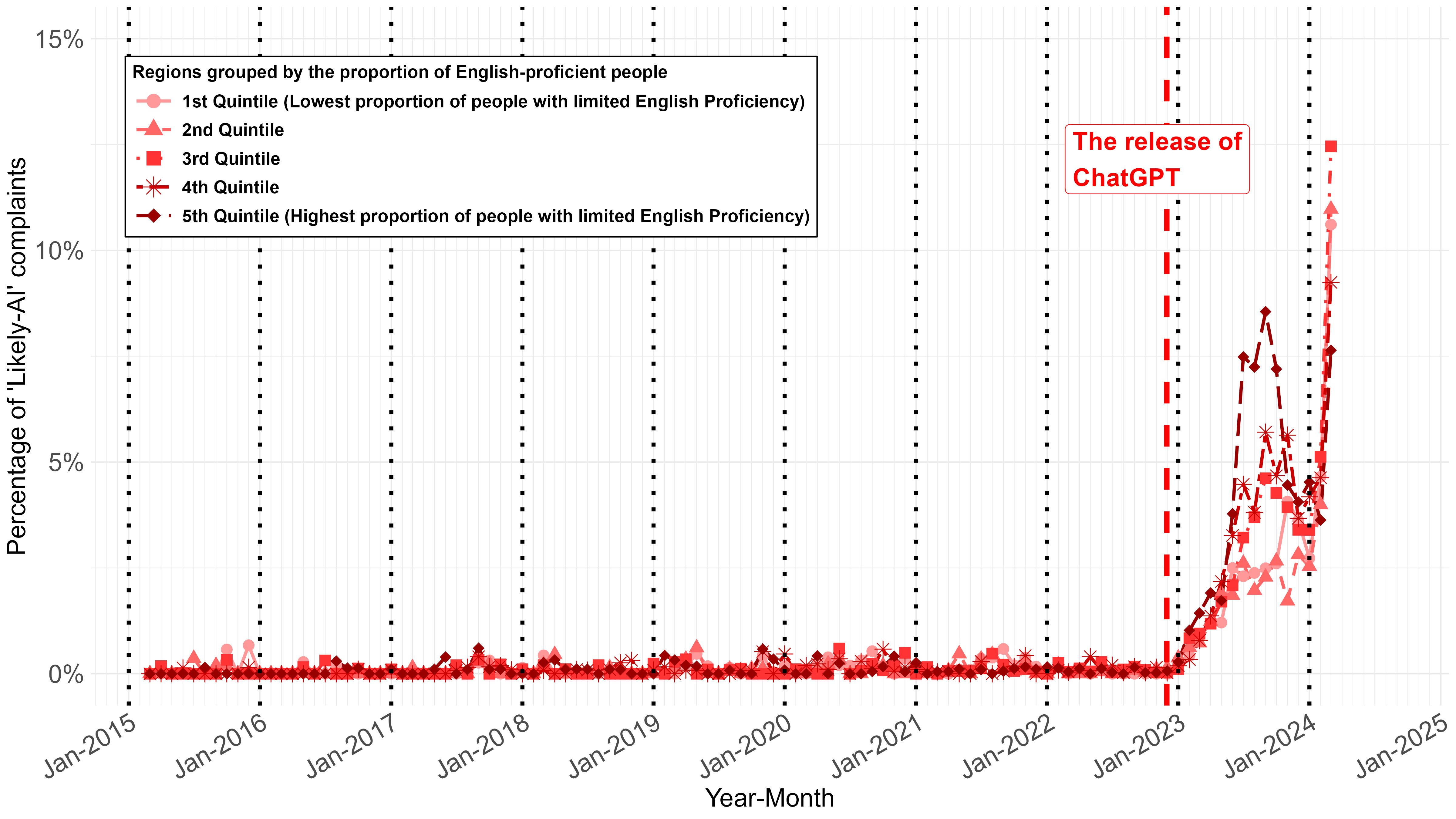}
         \caption{LLM Adoption Patterns Varying Across Regions Based on English proficiency}
    \label{fig:hetero_adoption_full}
\end{figure}
This figure displays regional variations in LLM adoption from 2015 through 2024, grouped by five English‐proficiency quintiles (based on 2017–2021 ACS data) at the ZIP‐code level. Each color–marker combination corresponds to a different quintile of ZIP codes (the deeper the red, the higher the proportion of individuals with limited English proficiency). For instance, the light pink line with circle markers (1st quintile) represents areas with the lowest average share of limited‐English‐proficient residents (around 4\%), while the bold red line with diamond markers (5th quintile) indicates the highest share (around 60\%). The figure demonstrates that early adoption of LLMs was more rapid in regions with a higher proportion of people with limited English proficiency.

\subsection{Additional Empirical Evidence} 
\label{si_section:additional_empirical}
% Our findings on the causal effect are robust to different specifications of the two-stage estimation. In particular, the robustness of our second-stage results holds under various first-stage model choices, such as the linear probability model (LPM), often used even when the outcome variables are binary. The use of the LPM in the first stage is common practice although the endogenous variable (i.e., LLM usage) is binary. In Table \ref{tab:lpm_residual}, the first column reports the second-stage result using the residuals, $\hat{v}_i$, obtained from the first-stage LPM. It shows a positive and significant coefficient; furthermore, the larger effect size may be mainly driven by the first-stage LPM predictions, which can be much greater than 1 or much smaller than 0.

Here we provide additional results from different levels of analysis. First, we present a simple comparison of the probability of receiving relief between Likely-AI complaints and Likely-Human complaints, incorporating various control variables. This complaint-level analysis, while controlling for many factors, does not fully address potential confounding variables and thus indicates association rather than causation. Next, we offer an analysis comparing different regions with varying proportions of Likely-AI complaints. We examine how these regions differentially improve in obtaining relief, employing a method similar to the canonical difference-in-differences analysis.

\paragraph{Simple Logistic Regression} Table \ref{tab:simple_reg} illustrates that the positive association between LLM usage and the likelihood of receiving relief remains statistically significant across various control specifications. Column (1) presents a baseline comparison between Likely-AI and Likely-Human complaints, showing a strong positive relationship. Column (2) shows that this relationship remains positive and statistically significant even after including time and ZIP code fixed effects, suggesting that broad temporal dynamics and ZIP-code-level factors cannot fully explain the difference. Finally, Column (3) incorporates complaint-issue fixed effects, demonstrating that the association persists net of potentially confounding differences in complaint topics.

\begin{table}[!h]
\vspace{0.3cm}
\footnotesize
\centering 
\caption{Simple Comparison with Various Control Variables}
\label{tab:simple_reg}
\begin{tabularx}{0.68\textwidth}{p{4.3cm} >{\centering\arraybackslash}p{1.7cm} >{\centering\arraybackslash}p{1.7cm} >{\centering\arraybackslash}p{1.7cm}}
\toprule
& \multicolumn{3}{c}{\textit{Dependent variable:}} \\ 
& \multicolumn{3}{c}{\textit{Relief Dummy}} \\
\cmidrule(lr){2-4}
 & (1) &  (2) & (3) \\ 
\midrule
Constant & \hspace{0.1cm}$-$0.408$^{***}$ &    &  \\ 
 & (0.004) &    &  \\ 
Likely-AI Dummy & \hspace{0.35cm}0.381$^{***}$ & \hspace{0.35cm}0.225$^{***}$   & \hspace{0.35cm}0.199$^{***}$ \\ 
 & (0.021) & (0.023)   &  (0.024)\\ 
\midrule
Year-Month Fixed Effects & N &  Y & Y \\
ZIP-code Fixed Effects & N &  Y & Y \\
Complaint Issue Fixed Effects & N &  N & Y \\
Observations & 244,407 & 244,407 & 244,407\\ 
Unit of Observations & \multicolumn{3}{c}{Each Complaint} \\
Sample Period & \multicolumn{3}{c}{Dec-2022 to Mar-2024}\\ 
\bottomrule
\multicolumn{4}{r}{\textit{Note.} $^{**}$ \textit{p} $<$ .01; $^{***}$ \textit{p} $<$ .001}
\end{tabularx}
\end{table}

\paragraph{Robustness check} Table \ref{tab:robust_more_fe} presents results from our two-stage regressions with company fixed effects. The primary finding that consumer use of LLMs in drafting complaints can increase the likelihood of receiving relief remains robust with company fixed effects.

Among more than 2,000 unique companies in our sample, complaints are heavily concentrated among the top 200 firms, which account for 95\% of all complaints. Because smaller firms often receive only a few complaints (typically from a narrow set of zip codes over short intervals), they provide insufficient data variation for controlling for other fixed effects. Restricting the analysis to the top 200 firms ensures an adequate number of observations per firm. Accordingly, we report the results based on this subsample here.

Including company fixed effects in both the AI-usage (first-stage) and relief-decision (second-stage) regressions eliminates time-invariant, firm-specific heterogeneity (e.g., distinct corporate cultures or customer-service policies) that might otherwise confound the estimates. Effectively, we compare each firm to itself over time, ruling out the possibility that certain companies are simply more prone to receive AI-generated complaints or more inclined to grant relief.

\begin{table}[!htp]
\centering
\footnotesize
\caption{Two-Stage Regression Results with Company Fixed Effects (Top 200 Companies Accounting for 95\% of the Complaints in Our Sample)}

%---------------------------%
% Panel A: First Stage
%---------------------------%
%-1.258e+00  2.550e-01 
%2.036e-01  6.595e-02  
\textbf{Panel A: First-Stage Regression Results} \\[4pt]
\begin{tabular}{@{\extracolsep{4pt}}lccc}
\toprule
 & \multicolumn{1}{c}{\textit{Dependent variable: Likely-AI Dummy}} \\
\midrule
Proportion of Households without Internet access 
  & $-$1.258$^{***}$ \\ 
  & (0.255)          \\[4pt]
Proportion of People with less English proficiency 
  & \phantom{x}0.204$^{**}$ \\ 
  & (0.066)       \\
\midrule
Year-Month Fixed Effect       & Y  \\
State Fixed Effect            & Y  \\
Complaint Issue Fixed Effect  & Y  \\
Finance Product Fixed Effect  & Y  \\
Company Fixed Effect & Y  \\ 
Demographic variables Included & Y  \\ 
Observations                  & \multicolumn{1}{c}{232,308} \\
Unit of Observations          & \multicolumn{1}{c}{Each Complaint} \\
Sample Period                 & \multicolumn{1}{c}{Dec-2022 to Mar-2024} \\ 
\bottomrule
\end{tabular}

\vspace{2em} % Add some vertical space between panels

%---------------------------%
% Panel B: Second Stage
%---------------------------%
\textbf{Panel B: Second-Stage Regression Result} \\[4pt]
\begin{tabular}{@{\extracolsep{4pt}}lccc}
\toprule
 & \multicolumn{1}{c}{\textit{Dependent variable: Relief Dummy}} \\ 
 %1.875e-01  4.426e-02 
 %1.880e-01  4.436e-02
 %1.878e-01  4.428e-02 
\midrule
Likely-AI Dummy 
  & \phantom{x} 0.188$^{***}$ \\ 
  & (0.044)       \\
\midrule
First-Stage IV &  Both IVs\\
Year-Month Fixed Effect      & Y  \\
State Fixed Effect           & Y  \\
Complaint Issue Fixed Effect & Y  \\
Finance Product Fixed Effect & Y  \\
Company Fixed Effect & Y  \\ 
Demographic variables Included & Y  \\ 
Observations                 & \multicolumn{1}{c}{232,308} \\
Unit of Observations         & \multicolumn{1}{c}{Each Complaint} \\
Sample Period                & \multicolumn{1}{c}{Dec-2022 to Mar-2024} \\
\bottomrule
\multicolumn{2}{r}{$^{*}$p$<$0.05; $^{**}$p$<$0.01; $^{***}$p$<$0.001} 
\end{tabular}

\label{tab:robust_more_fe}
\end{table}

Table \ref{tab:robust_resid} demonstrates that our core finding, the relationship between LLM usage and the likelihood of obtaining relief, remains robust under multiple specifications. In Columns (1) and (2), we vary how the first‐stage residual is constructed, echoing prior findings that different residual formulas can yield differing effect sizes \cite{garrido2012choosing}. Although we observe some variability in magnitude, the coefficient on the Likely‐AI dummy remains statistically significant throughout. For the reasons detailed in \cite{garrido2012choosing,basu20182sls}, we ultimately focus on the standardized residual, as raw residuals may be problematic in certain contexts. In Column(3), we replace the logit‐based model with a linear probability model. Although predicted values from the linear model can lie outside the unit interval, the linear model is often employed as an approximation for causal inference \cite{angrist2009mostly}, and we again find that LLM usage significantly increases the probability of receiving relief. Across all specifications, the magnitude of this effect is consistently positive and statistically very significant, underscoring the robustness of our primary conclusion.

\begin{table}[!ht]
\vspace{0.3cm}
\footnotesize
\centering 
\caption{Robustness check regarding the choices of residual formulas and models}
\label{tab:robust_resid}
\begin{tabularx}{0.88\textwidth}{p{5.3cm} >{\centering\arraybackslash}p{2.4cm} >{\centering\arraybackslash}p{2.4cm} >{\centering\arraybackslash}p{2.4cm}}
\toprule
& \multicolumn{3}{c}{\textit{Dependent variable:}} \\ 
& \multicolumn{3}{c}{\textit{Relief Dummy}} \\
\cmidrule(lr){2-4}
 & (1) &  (2) & (3) \\ 
\midrule
Likely-AI Dummy & \hspace{0.35cm}1.315$^{***}$ & \hspace{0.35cm}1.992$^{***}$   & \hspace{0.35cm}0.296$^{***}$ \\ 
 & (0.184) & (0.149)   &  (0.032)\\ 

 % 1.992e+00  1.487e-01
 %  2.956e-01  3.216e-02
\midrule
 Residual Formula  & raw residual   & deviance residual & raw residual \\
    Model (First Stage \& Second Stage)             & Logit \& Logit & Logit \& Logit & Linear  \& Linear  \\
    Instruments & Both & Both & Both \\
Year-Month Fixed Effects & Y &  Y & Y \\
ZIP-code Fixed Effects & Y &  Y & Y \\
Complaint Issue Fixed Effects & Y &  Y & Y \\
Observations & 244,407 & 244,407 & 244,407\\ 
Unit of Observations & \multicolumn{3}{c}{Each Complaint} \\
Sample Period & \multicolumn{3}{c}{Dec-2022 to Mar-2024}\\ 
\bottomrule
\multicolumn{4}{r}{\textit{Note.} $^{**}$ \textit{p} $<$ .01; $^{***}$ \textit{p} $<$ .001}
\end{tabularx}
\end{table}

\paragraph{Diff-in-Diff Analysis} To further estimate the impact of LLM usage on relief decisions, we use complaints submitted both before and after the release of ChatGPT, and employ a canonical difference-in-differences approach, comparing regions where consumers have begun adopting LLMs (treatment group) with those where they have not (control group). Specifically, we estimate the following regression model:
\vspace{-0.1cm}
\begin{equation*}
Y_{jt} = \beta_1 (D_j \times Post_t)  + \alpha_j + \gamma_t + \varepsilon_{jt}
\label{eq:did}
\end{equation*}
\vspace{-1cm}

where $Y_{jt}$ is the proportion of complaints receiving relief from firms in ZIP code $j$ at time $t$; $D_j$ is an indicator equal to 1 if ZIP code $j$ has any complaint likely written by LLMs after the release of ChatGPT, and 0 otherwise; $Post_t$ is an indicator equal to 1 for periods after November 30, 2022 (i.e., the release date of ChatGPT), and 0 otherwise; $\alpha_j$ are ZIP code fixed effects; $\gamma_t$ are year-month fixed effects; and $\varepsilon_{jt}$ is the error term. The coefficient $\beta_1$ captures the average treatment effect of LLM usage on the proportion of complaints receiving relief. Standard errors are clustered at the ZIP code level to account for serial correlation within ZIP codes over time.

In addition to our primary specification, we explore several alternative models, as reported in Table \ref{tab:did}. In Column (1), we define the treatment group as ZIP codes that recorded at least one Likely-AI complaint following the release of ChatGPT, while the control group comprises ZIP codes that did not report any such complaints during the same period. In Columns (2) and (3), we modify the standard Difference-in-Differences framework by replacing the interaction term $D_j \times Post_t$ with a time-varying treatment intensity measure, $D_{jt}$, to account for varying Likely-AI complaint patterns within individual ZIP codes across different time periods. Specifically, in Column (2), $D_{jt}$ is a time-varying binary variable indicating whether a ZIP code has any Likely-AI complaints in a given period. In Column (3), $D_{jt}$ represents the proportion of Likely-AI complaints in each ZIP code by period. This adjustment acknowledges that even within treated ZIP codes, there may be periods in the post-treatment phase where no Likely-AI complaints are recorded, thus capturing heterogeneity in treatment intensity over time.

\begin{table}[!h]
\vspace{0.3cm}
\footnotesize
\centering 
\caption{Difference-in-Differences Estimation Results}
\label{tab:did}
\begin{tabularx}{0.68\textwidth}{p{4.3cm} >{\centering\arraybackslash}p{1.7cm} >{\centering\arraybackslash}p{1.7cm} >{\centering\arraybackslash}p{1.7cm}}
\toprule
& \multicolumn{3}{c}{\textit{Dependent variable:}} \\ 
& \multicolumn{3}{c}{\textit{Proportion of complaints with relief}} \\
\cmidrule(lr){2-4}
 & (1) &  (2) & (3) \\ 
\midrule
$D_j \times Post_t$ & \hspace{0.35cm}0.057$^{***}$ &    &  \\ 
 & (0.003) &    &  \\ 
$D_{jt}$ &  &  \hspace{0.35cm}0.075$^{***}$ & \hspace{0.35cm}0.104$^{***}$ \\ 
 &  &   (0.004) & (0.012) \\ 
\midrule
Year-Month Fixed Effects & Y &  Y & Y \\
ZIP Code Fixed Effects & Y &  Y & Y \\
Observations & 350,824 &  350,824 & 350,824 \\ 
Unit of Observations & \multicolumn{3}{c}{ZIP Code ($j$) and Year-Month ($t$)} \\
Sample Period & \multicolumn{3}{c}{Mar-2015 $\sim$ Mar-2024} \\ 
\bottomrule
\multicolumn{4}{r}{\textit{Note.} $^{**}$ \textit{p} $<$ .01; $^{***}$ \textit{p} $<$ .001}
\end{tabularx}
\end{table}

The estimation results presented in Table \ref{tab:did} suggest a statistically significant increase in the probability of obtaining relief from firms in regions where consumers adopted LLMs. Specifically, the likelihood of receiving relief in these regions is estimated to be between 5.7 and 10.4 percentage points higher relative to regions where consumers did not utilize LLMs in drafting their complaints.

% \hl{Addising discussion, what we can achieve by analyzing the inidividual level,  not regional level -- need to provide a stronger justification why we are doing what we are doing in our main analysis.}

\section{Pilot Studies and Development of Stimuli for the Main Experiments}
\label{si_section:pilot_studies_and_stim_dev}

\subsection{Pilot Study 1}
\label{si_subsection:pilot_study_1}
Pilot Study 1 was conducted to validate the participant pool used for the main experiments. Specifically, Pilot Study 1 tested whether participants recruited from Prolific would behave like the decision-makers at financial firms that actually responded to consumer complaints, as recorded in the CFPB data.

\paragraph{Stimuli} We selected consumer complaints from the CFPB data to use as stimuli for Pilot Study 1 by taking the following steps. First, from the CFPB data set containing 1.13 million complaints, we excluded complaints that were written after December 31, 2021, which left us with a set of 549,908 consumer complaints; this step ensured that our stimuli would consist only of consumer complaints that were written before the release of popular LLMs (e.g., the release of ChatGPT in November 2022). Second, we then excluded complaints for which the financial company's response was neither ``Closed with monetary relief'' nor ``Closed with explanation,'' which left us with a set of 489,582 complaints; this step ensured that our stimuli would consist only of complaints that either led to actual monetary relief from a financial firm or not. Third, we further narrowed our set of complaints to those concerning the top three most common financial products (``Credit reporting, credit repair services, or other personal computer reports,'' ``Debt collection,'' and ``Mortgage''); this left us with a set of 318,109 complaints. Fourth, for each of the top three products, we identified the top three most common issues and created a subset of complaints for each of the three issues, for a total of nine subsets of complaints (3 most common products $\times$ 3 most common issues). Examples of the issues included ``Attempts to collect debt not owed'' for the ``Debt collection'' product and ``Struggling to pay mortgage'' for the ``Mortgage'' product; the full lists of issues can be found in the CSV files uploaded on the \textcolor{blue}{\href{https://osf.io/fh2cz/?view_only=0d1f07d0baf54cfdb8aa6fb30e02f9b3}{OSF page}}. Fifth, we divided each of the nine subsets of complaints into two smaller subsets of complaints: those that led to actual monetary relief and those that did not; this step created a total of 18 subsets of complaints. Sixth, from each of the 18 subsets of complaints, we randomly selected 18 complaints (hereafter, the \textit{final 18 subsets of complaints}) and sorted the selected complaints from the shortest to the longest in length. Lastly, within each of the final 18 subsets of complaints, we reviewed the complaints from the top (i.e., the shortest complaint) and selected six complaints that were appropriate to include as the stimuli for Pilot Study 1 (e.g., complaints that did not feature too much redacted text). This resulted in a total of 108 consumer complaints that would compose the stimuli for Pilot Study 1 (6 complaints within each of the 18 final subsets of complaints).

\paragraph{Method} We recruited 216 participants ($\textit{M}_{age}$ = 36; 48\% men, 50\% women, and 2\% other) from Prolific. Participants were thanked for participating in the study and were informed that the researchers were interested in learning about ``how consumer complaints might be handled by financial institutions.'' Participants were told that they would see a total of five different complaints and were asked to imagine that they were ``in charge of addressing [each] complaint.'' Participants were then presented with a series of five complaints, which were randomly selected from the set of 108 complaints (see the Stimuli section above). For each complaint, they were asked the main dependent measure of the study: ``If you were handling the complaint above, how likely would you be to offer monetary compensation to address the consumer's complaint?'' Participants answered the question on a 7-point scale (1 = \textit{Not likely at all}, 7 = \textit{Extremely likely}). After reporting their likelihood of providing hypothetical compensation for five complaints, participants reported their age and gender to complete the study.

\paragraph{Results} Aggregating data across all five ratings and across all participants, we first conducted a \textit{t}-test with hypothetical monetary compensation rating as the dependent variable and whether a complaint had resulted in actual monetary relief or not as the independent variable\footnote{Technically, this independent \textit{t}-test is not appropriate, because subsets of the compensation likelihood ratings come from the same entities (individual participants), and therefore not all observations are independent from one another. However, we include the results from this test for a more intuitive understanding of our results. These results align very closely with results from the appropriate analyses (linear fixed effects models), which are discussed immediately afterwards.}. Not surprisingly, participants were more likely to offer hypothetical monetary compensation for the complaints that had resulted in actual monetary relief from firms (\textit{M} = 4.75, \textit{SD} = 1.84) than for those that had not resulted in actual monetary relief from firms (\textit{M} = 4.20, \textit{SD} = 1.98), \textit{t}(1078) = 4.74, \textit{p} \textless~.001. More importantly, we find the same pattern of results in preregistered linear fixed effects models, as shown in Table \ref{tab:pilot_study_1_results}. That is, across various linear fixed models that control for the effects of products, issues, and participants, participants' decisions to offer hypothetical monetary compensation was positively related to whether complaints had previously resulted in actual monetary relief from firms (\textit{b}s > 0.52, \textit{p}s < .001). In other words, participants recruited from Prolific were more likely to offer hypothetical monetary compensation for the complaints that had actually resulted in relief from firms than for those that had not actually resulted in relief, demonstrating an ability to evaluate complaints similar to that of the decision-makers at financial firms.

\begin{table}[!ht] 
\centering 
\footnotesize
\caption{Results From Pilot Study 1}\label{tab:pilot_study_1_results} 
\resizebox{\textwidth}{!}{%
% \begin{tabular}{@{}p{5.5cm}ccccccc@{}}
\begin{tabular}{@{\extracolsep{5pt}}lccccc}
\\[-1.8ex]\hline 
\hline \\[-1.8ex] 
 & \multicolumn{5}{c}{\textit{Dependent variable: Compensation likelihood}} \\ 
\cline{2-6} 
\\[-1.8ex] & \multicolumn{5}{c}{Model} \\ 
\\[-1.8ex] & 1 & 2 & 3 & 4 & 5 \\ 
\hline \\[-1.8ex] 
Constant 
    & 4.198$^{***}$ &  &  &  &   \\ 
    & (0.0823) &  &  &  &  \\ 
Actual Monetary Relief From Firm (No vs. Yes) 
    & 0.552$^{***}$ & 0.552$^{***}$ & 0.552$^{***}$ & 0.532$^{***}$ & 0.525$^{***}$ \\ 
    & (0.116) & (0.115) & (0.116) & (0.104) & (0.102)\\ 
\hline \\[-1.8ex] 
Product Fixed Effects               & N & Y & N & N & Y \\ 
Issue Fixed Effects                 & N & N & Y & N & Y \\ 
Participant Fixed Effects           & N & N & N & Y & Y \\ 
\hline 
\hline \\[-1.8ex] 
\multicolumn{6}{p{\textwidth}}{\textit{Note.} $^{***}$\textit{p} $<$ .001. \textit{N} = 1080 (5 ratings $\times$ 216 participants).} \\ 
\end{tabular} 
}
\end{table}

\subsection{Pilot Study 2}
\label{si_subsection:pilot_study_2}
Pilot Study 2 was conducted to determine which linguistic features people would be more likely to focus on when editing complaints with an LLM.

\paragraph{Method} We recruited 212 participants ($\textit{M}_{age}$ = 40; 49\% men, 49\% women, and 2\% other) from Prolific. Participants began the study by learning that consumers can report complaints regarding their financial institutions to the CFPB and that such complaints can later be accessed by the public. Participants were then told that they would see five different complaints and that for each complaint, their task was to (1) imagine that they wrote the complaint to ``obtain monetary relief from a financial institution'' and (2) indicate how they would use ChatGPT in editing their complaint ``to improve [their] chances of obtaining monetary compensation.''

Participants then proceeded to review a set of five random complaints from a set of 150 complaints selected from the CFPB dataset. 
% (details on selecting these 150 complaints are provided in the next section)
Participants read one complaint at a time, and for each complaint, they answered the study's main dependent measure, consisting of 10 items, each with a 5-point scale of disagreement or agreement: ``For this complaint, I would ask ChatGPT to make the complaint...(1) more polite, (2) more coherent, (3) more fact-based, (4) more persuasive, (5) more clear, (6) more professional, (7) more assertive, (8) more concise, (9) less repetitive, and (10) less emotional (1 = \textit{Strongly Disagree}, 5 = \textit{Strongly Agree}).'' (We selected these 10 linguistic features loosely based on our conversation with ChatGPT [e.g., \href{https://chat.openai.com/share/f1c6e563-3b98-4364-9315-f87d62103dda}{https://chat.openai.com/share/f1c6e563-3b98-4364-9315-f87d62103dda}] and a discussion between the authors.) After completing the 10-item dependent measure for each of the five complaints, participants answered an open-ended question, ``If you were to use ChatGPT to write (draft or edit) a complaint regarding an issue with your financial institution, what kind of requests would you make to ChatGPT?'' Participants then provided demographic information (age, gender, race or ethnicity, education, household income, and political ideology) to complete the study.

\paragraph{Results} Results showed that the three features that participants expressed greatest intent to focus on were (1) clarity, (2) coherence, and (3) professionalism (see Table \ref{tab:top_3_ling_features}). We thus centered our investigation on these three linguistic features as linguistic features that people are likely to focus on when editing their complaints with LLMs.
% See \textcolor{blue}{\textit{Online Appendix}, section \ref{si_subsection:pilot_study_procedure}} for more details on this pilot study.

\hfill % Adds horizontal space between the figure and table
\begin{center} % Centers the minipage on the page
    \begin{minipage}[c]{1\textwidth} % Adjust width to suit your needs
        \centering
        \fontsize{9pt}{10pt}\selectfont
        \setlength{\tabcolsep}{4pt} % Optional: reduces the space between columns
        \renewcommand{\arraystretch}{0.8} % Reduces the vertical space between rows
        \begin{tabular}{@{\extracolsep{5pt}}lcc} 
        
            \Xhline{2\arrayrulewidth}
            \textbf{Linguistic Feature} & \makecell{\textbf{Intent to Focus} \\[-1.5ex] \textbf{Rating Mean [95\% CI]}} \\
            \hline
            More clear & 3.78 [3.71, 3.86]  \\
            More coherent & 3.71 [3.64, 3.79]  \\
            More professional & 3.71 [3.63, 3.79]  \\
            More concise & 3.62 [3.54, 3.70]  \\
            More persuasive & 3.54 [3.47, 3.61]  \\
            More fact-based & 3.21 [3.13, 3.29]  \\
            More assertive & 3.05 [2.97, 3.13]  \\
            More polite & 2.86 [2.79, 2.94]  \\
            Less repetitive & 2.82 [2.74, 2.90]  \\
            Less emotional & 2.82 [2.74, 2.90]  \\
            \Xhline{2\arrayrulewidth}
        \end{tabular}
        \captionof{table}{Participant Ratings on the Intent to Modify Linguistic Features}
        \label{tab:top_3_ling_features}
            
    \end{minipage}
\end{center}
\vspace{1cm}

\subsection{Stimuli Development}
\label{si_subsection:stimuli_development}

\begin{figure}[!ht]
    \centering
    \includegraphics[width=10cm]{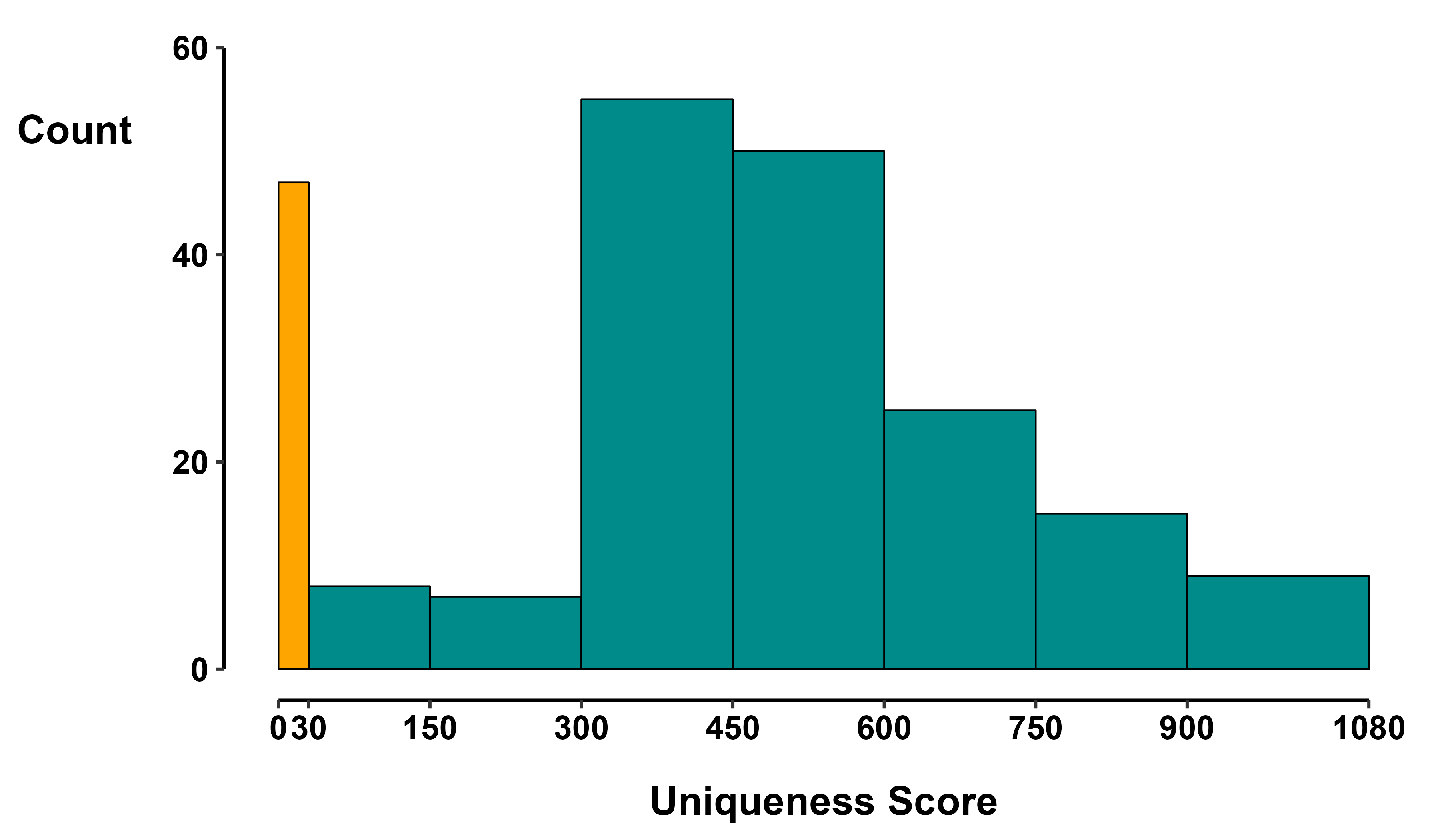}
    \caption{This figure shows the counts of consumer complaints for given ranges of uniqueness scores for the 216 nonduplicated complaints that the AI detection tool determined as completely written by humans. We randomly selected 20 complaints from the set of complaints with uniqueness scores larger than 30.}
    \label{fig:uniqueness_score}
    \vspace{-0.15cm}
\end{figure}

As mentioned in the main text, the stimuli used in the main experiments were 80 consumer complaints, 20 of which were the original, unedited complaints from the CFPB data set (\textit{Control} complaints), and 60 of which were three different edited versions of these 20 complaints~\textemdash~namely, 20 \textit{Clear}, 20 \textit{Coherent}, and 20 \textit{Professional} complaints. These stimuli were carefully developed in three stages with the following steps:
\\\\\textbf{Stage 1: Selection of Complaints for a Pilot Study}
\begin{enumerate}[itemsep=0pt, parsep=0pt]
\item From the CFPB dataset, we identified consumer complaints that the AI detection tool identified as completely written by humans (i.e., those with the AI scores of 0).
\item From this set of \textit{Likely-Human} complaints, we randomly selected a set of 300 consumer complaints.
\item From the above set of 300 complaints, we eliminated 84 complaints that were exact duplicates of other complaints\footnote{Interestingly, we encountered duplicated complaints that may have been written by legal agents representing their clients (e.g., those that cited precise legal code like ``Fair Credit Reporting Act - 15 U.S.C.1681 section 602''). Investigating the potential influence of legal representatives (i.e., the effect of ``lawyering up'' on persuasive messages) could be an interesting direction for future research. However, more importantly for our purposes, inclusion or exclusion of duplicated messages (including such messages possibly written by legal representatives) did not change our results. Specifically, reanalyses of the data after excluding the duplicated messages do not qualitatively change the direction nor statistical significance of the results from our observational studies.\label{footnote:legal_experts}}.
\item Upon examining the remaining 216 complaints, we observed groups of very similar complaints. To avoid presenting similar complaints to the experiment participants and to represent a diverse range of issues in consumer complaints, we sought to select a set of unique complaints. To do so, we established a uniqueness score for each complaint based on Levenshtein distance \cite{levenshtein1966binary}, a measure of dissimilarity between two texts.
\item Specifically, for each of the 216 complaints, we calculated its Levenshtein distance with every other complaint in the set, recording the smallest of these distances as the complaint's \textit{uniqueness score}. For instance, to determine the first complaint's uniqueness score, we calculated the Levenshtein distance between it and each of the remaining 215 complaints; then, we took the minimum of the 215 Levenshtein distances and recorded this value as the first complaint's uniqueness score. A high uniqueness score (i.e., a high minimum Levenshtein score) would mean that the given complaint is highly different from the complaint that is most similar to it, and even more different from all the other complaints. Figure \ref{fig:uniqueness_score} of this document shows a histogram of the uniqueness scores for all 216 complaints.
\item After examining the uniqueness scores, we discarded complaints with uniqueness scores of 30 or lower, as they were relatively similar to others. This reduced the set of complaints to 169.
\item From the set of 169 complaints, we randomly selected 150 complaints to use in a pilot study.
\end{enumerate}
\textbf{Stage 2: Pilot Study to Determine the Focal Linsguistic Features of the Main Experiment}
\begin{enumerate}[itemsep=0pt, parsep=0pt]
\setcounter{enumi}{7}
\item In the pilot study (\textit{N} = 212; described in greater detail in a subsequent section), we presented each participant with five complaints randomly chosen from the set of 150 complaints. For each of these complaints, we asked the given participant how they would use ChatGPT to edit their hypothetical complaints to obtain monetary relief from financial institutions. The results suggested that people would be likely to use ChatGPT to edit their complaints to be \textit{more clear}, \textit{coherent}, and \textit{professional}. We thus concluded that clarity, coherence, and professionalism are three linguistic features that people are relatively more likely to focus on when using an LLM like ChatGPT to edit persuasive messages. Accordingly, we focused on these three linguistic features in the main experiments. %See \textcolor{blue}{\textit{Online Appendix}, section 3} for details on this pilot study.
\end{enumerate}
\textbf{Stage 3: Final Selection of Complaints for the Main Experiment}

\begin{figure}[!ht]
    \centering
    \includegraphics[width=10.6cm]{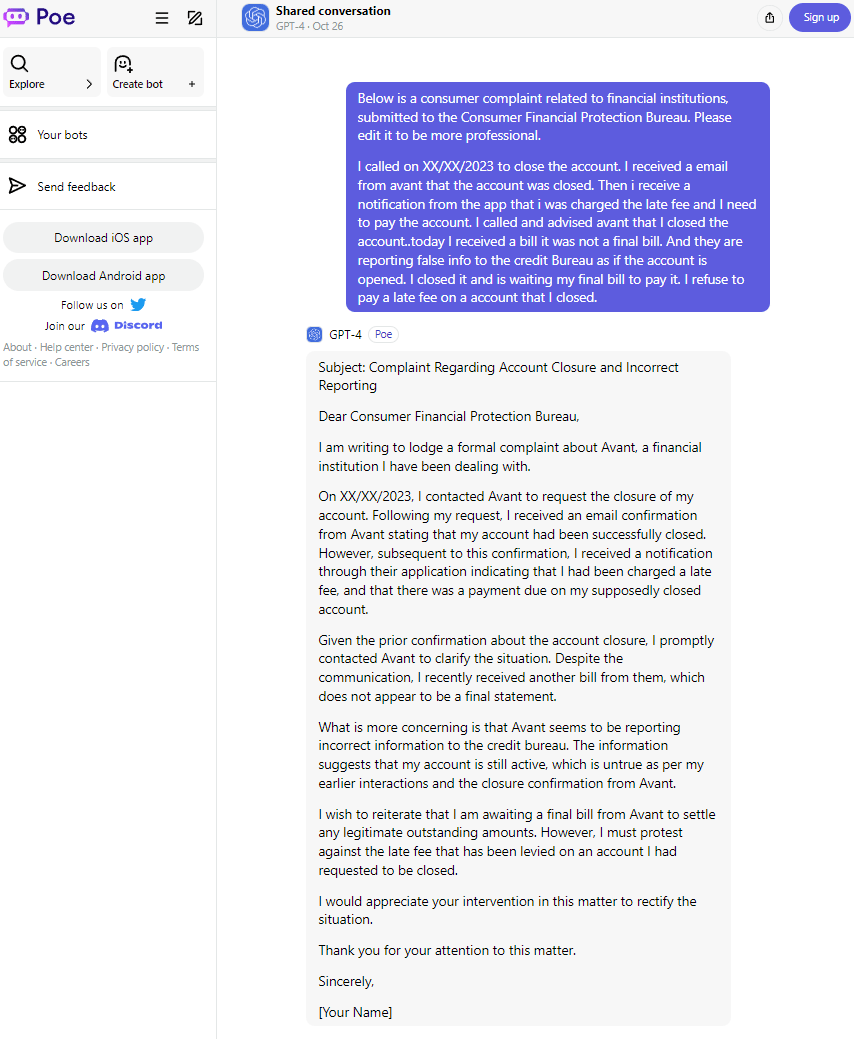}
    \caption{The figure shows the process through which the 60 edited complaints were generated. Specifically, we took the 20 unedited complaints written by actual consumers (from the CFPB data), presented each complaint to ChatGPT (one complaint at a time), and asked it to edit the complaint to be more clear, coherent, and professional as shown in the figure. The order of the three instructions (i.e., ``...edit it to be more clear / coherent / professional'') was randomized for each of the 20 complaints.}
    \label{fig:editing_with_chatgpt_example}
    \vspace{-0.15cm}
\end{figure}

\begin{enumerate}[itemsep=0pt, parsep=0pt]
\setcounter{enumi}{8}
\item We proceeded to select consumer complaints for the main experiments by returning to the set of 150 complaints (see Step 7 above). We sorted 150 complaints by their length (i.e., number of characters in complaint) and selected the 20 shortest complaints while excluding complaints that seem to have been written by legal experts (because we were more interested in how laypeople, rather than legal experts, would use ChatGPT to edit complaints; see Footnote \ref{footnote:legal_experts} of this document). We selected the shortest complaints to minimize the possibility that the main experiments' participants could be overwhelmed or disengaged by reading long complaints. The 20 complaints thus selected were used in the main experiments as the 20 unedited complaints (i.e., \textit{Control} complaints).
\item We proceeded to create edited versions of the 20 unedited (\textit{Control}) complaints. Specifically, we gave each of these 20 complaints to ChatGPT one at a time and asked ChatGPT to edit each complaint to be (1) more clear, (2) more coherent, and (3) more professional (with the order of these instructions randomized; see Figure \ref{fig:editing_with_chatgpt_example} for an example of one such instruction).
\item This resulted in a final set of 80 complaints in total, consisting of 20 unedited complaints (which we will refer to as Control complaints) and 60 edited complaints (20 Clear, 20 Coherent, and 20 Professional complaints).
\end{enumerate}

\vspace{2cm}

\clearpage

\clearpage
\newpage

\subsection{Pilot Study 3}
\label{si_subsection:pilot_study_3}
Pilot Study 3 tested whether the 60 LLM-edited complaints indeed showed enhancement in presentation relative to the 20 unedited complaints across the three linguistic features (clarity, coherence, and professionalism).

\paragraph{Method} We recruited 490 participants ($\textit{M}_{age}$ = 39; 49\% men, 50\% women, and 0.6\% other) from Prolific and randomly assigned them to one of three conditions in a between-subjects design: Clarity, Coherence, and Professionalism. In all conditions, participants were informed that they would be evaluating five consumer complaints in the finance industry. Participants were then presented with a series of five complaints, randomly drawn from the pool of 80 complaints (i.e., complaints from Section \ref{si_subsection:stimuli_development} above). We ensured that none of the five complaints featured the same content (see the procedure for Experiment 1). Participants rated each of the five complaints on the linguistic feature corresponding to their assigned condition: clarity, coherence, or professionalism. Specifically, for each complaint, they answered the question, ``How clear / coherent / professional is the complaint above?'' on a 7-point scale (1 = \textit{Not clear / coherent / professional at all}, 7 = \textit{Extremely clear / coherent / professional}). Participants then reported any issues on the survey (optional) and reported age and gender to complete the study.

\paragraph{Results} As expected, the 60 edited complaints showed enhancement in presentation compared to the 20 unedited complaints; see Panels A-C of Figure \ref{fig:pilot_study_results} in the main text. Interestingly, the edited complaints showed enhancement on \textit{all three} linguistic features measured. Although ratings of clarity, coherence, and professionalism were all independent from one another due to the between-subjects design of this study, the edited complaints still showed similar enhancements across all three linguistic features. More importantly, results from Pilot Study 3 confirm that the 60 LLM-edited complaints were enhanced in presentation relative to the 20 unedited complaints, thus validating them as appropriate stimuli to be used in the main experiments.

\section{Controlled Experiments}
\label{si_section:experiments}

\subsection{Experiment 1}
\label{si_subsection:exp_1}

\paragraph{Method} We recruited 301 participants ($\textit{M}_{age}$ = 42; 49\% men, 50\% women, and 1.3\% other) from Prolific. Participants were thanked for participating in the study and were informed that the researchers were interested in learning about ``how consumer complaints might be handled by financial institutions.'' Participants were told that they would see a total of five different complaints and were asked to imagine that they were ``in charge of addressing the complaint.'' Participants were then presented with a series of five complaints, randomly chosen from the set of 80 complaints as follows:
\begin{enumerate}[itemsep=0pt, parsep=0pt]
\item A complaint from the pool of 20 unedited complaints (1 of 20 Control Complaints)
\item A complaint from the pool of 20 complaints edited with ChatGPT to be more clear (1 of 20 Clear Complaints)
\item A complaint from the pool of 20 complaints edited with ChatGPT to be more coherent (1 of 20 Coherent Complaints)
\item A complaint from the pool of 20 complaints edited with ChatGPT to be more professional (1 of 20 Professional Complaints)
\item An additional complaint randomly chosen the four types of complaints above
\end{enumerate}

We ensured that each of the five randomly chosen complaints featured different complaint content. That is, no two complaints were the same complaints or originated from the same unedited complaint. For each complaint, we asked participants to rate the likelihood of providing hypothetical monetary compensation regarding each complaint (``If you were handling the complaint above, how likely would you be to offer monetary compensation to address the consumer’s complaint? [1 = \textit{Not likely at all}, 7 = \textit{Extremely likely}]''). After making hypothetical compensation decisions for five complaints, participants reported any issues with the study (optional) and reported their age and gender to complete the study.

\begin{figure}[!h]
    
    \centering
    \caption{Compensation Likelihood for Unedited and LLM-Edited Complaints (Experiment 1 Results)}
    \includegraphics[width=\textwidth]{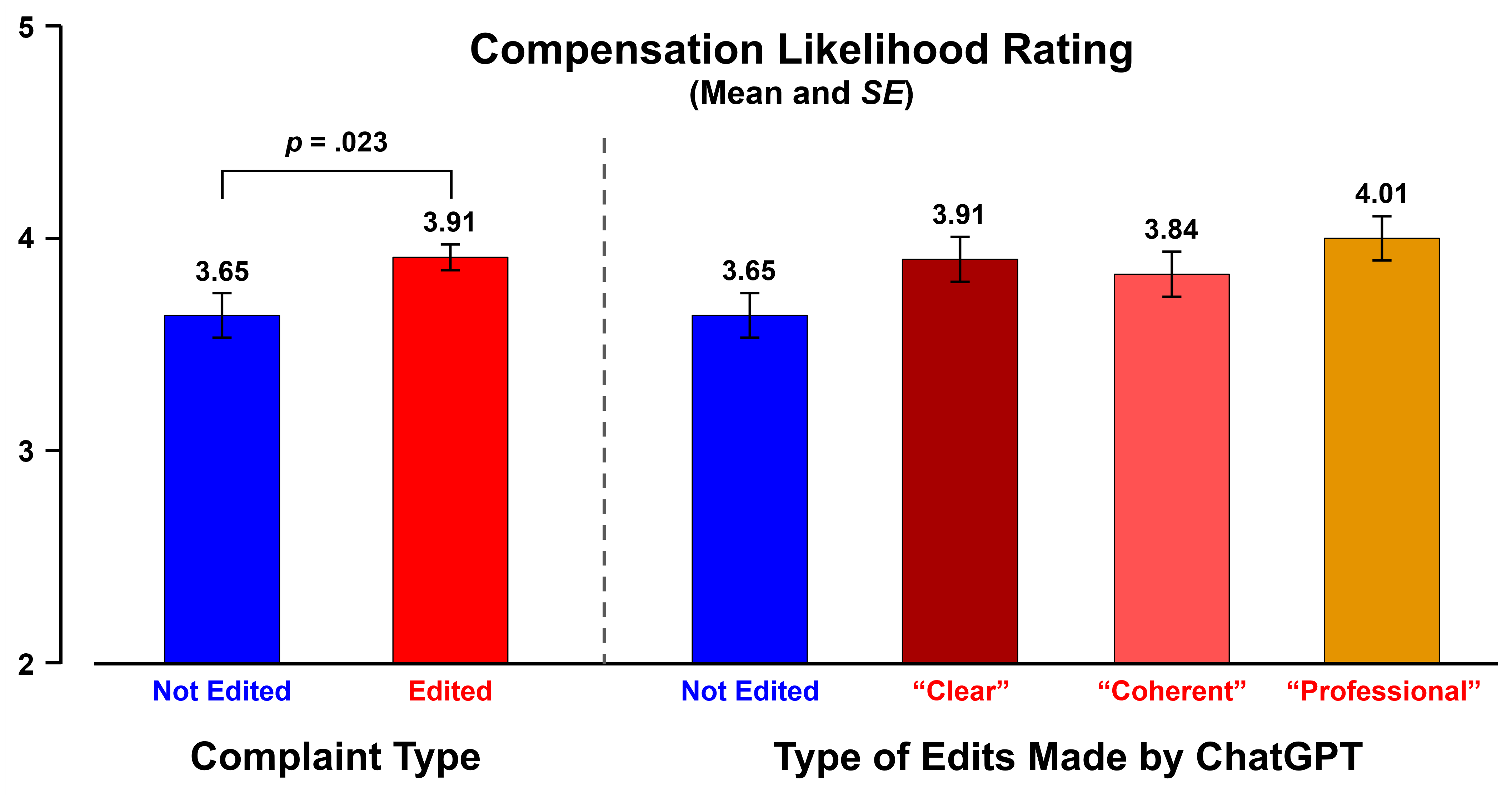}
    \begin{minipage}{0.98\linewidth}
    \vspace{1em}
    \small \textit{Note.} The figure shows the results from Experiment 1. The left portion shows the mean and \textit{SE} of the compensation ratings for the 20 unedited complaints (Control) and 60 edited complaints (Clear, Coherent, and Professional combined), while the right portion shows the same results separately for each type of LLM edits. 
    % For testing our preregistered hypotheses, we estimated the increase in compensation likelihood for edited complaints (Clear, Coherent, and Professional combined or separately) relative to Control complaints in linear fixed effects models, controlling for the fixed effects of individual participants and complaint contents. These estimates are reported in the main text and in the \textcolor{blue}{\href{https://osf.io/fh2cz/?view_only=0d1f07d0baf54cfdb8aa6fb30e02f9b3}{OSF page}}.    
    \end{minipage}    
    \label{fig:exp_1_results}
    \vspace{0.1cm}
\end{figure}

\paragraph{Results} Complaints edited with ChatGPT (\textit{Coherent}, \textit{Clear}, and \textit{Professional} complaints) were more likely to receive hypothetical monetary compensation (\textit{M} = 3.92, \textit{SD} = 2.04) than the unedited (\textit{Control}) complaints (\textit{M} = 3.65, \textit{SD} = 2.06), \textit{t}(1503) = 2.27, \textit{p} = .023, as illustrated in Figure \ref{fig:exp_1_results}. More importantly, the greater likelihood of compensation was also found in the preregistered linear fixed effects models controlling for the effects of complaint contents and participants; see Table \ref{tab:lfe_models_in_exp_1} in the main text. Specifically, the LLM-edited complaints were more likely to be offered hypothetical monetary compensation than the unedited complaints (\textit{b} = 0.31, \textit{SE} = 0.087, \textit{t} = 3.52, \textit{p} < .001). Thus, results from Experiment 1 directly support the finding from analyses of the CFPB data that editing complaints with an LLM significantly increases the likelihood of obtaining relief.

\subsection{Experiment 2}
\label{si_subsection:exp_2}

We conducted Experiment 2 with two goals in mind. First, we sought to replicate the findings from Experiment 1 with participants \textit{who have worked in the finance industry}. Second, we tested whether the previous results would be replicated also when participants focused on specific presentation features of complaints. 

Using Prolific, we recruited 400 individuals who have worked in the finance sector in the past ($\textit{M}_{age}$ = 45; 50.5\% men, 49\% women, and 0.5\% other). We began the study by giving the participants some ideas about the study’s purpose (``We’re interested in learning about how consumer complaints might be handled by financial institutions'') and informing them of why they were recruited (``We sought to recruit participants like you who... have past work experience in the finance sector''). Participants then learned that they would encounter four different consumer complaints and were asked to imagine that they were in charge of handling each of the complaints. The key manipulation in Experiment 2 was asking participants to focus on one linguistic feature\textemdash clarity or professionalism\textemdash and asking them to use it as the main judgment criterion to decide their likelihood of providing compensation. Specifically, for each complaint, we asked participants to rate it on the focal linguistic feature they were assigned (either clarity or professionalism) using the following 7-point scale: ``How clear [professional] is the complaint above? (1 = Not clear [professional] at all, 7 = Extremely clear [professional] ).'' Participants then rated how likely they would be to provide compensation for the complaint: ``If you were handling the complaint above, how likely would you be to offer monetary compensation to address the consumer’s complaint? (1 = Not likely at all, 7 = Extremely likely).'' Participants provided these two ratings for each of the four randomly assigned complaints. They then provided demographic information to complete the study.

Experiment 2 replicated the findings from Experiment 1. Specifically, the complaints edited with ChatGPT (\textit{Clear}, \textit{Coherent}, and \textit{Professional} complaints combined) were more likely to receive hypothetical monetary compensation than the unedited complaints (\textit{Control} complaints only) in a linear fixed effects model which controlled for fixed effects of participants and complaint contents, \textit{b} = 1.01, \textit{SE} = 0.084, \textit{t} = 12.10, \textit{p} \textless~.001; see Model 2 in Table \ref{tab:lfe_models_in_exp_2}. When we estimate the same model separately within each of the two conditions (Focus on Clarity or Focus on Professionalism), we obtain similar results; see Models 5 and 6 in Table \ref{tab:lfe_models_in_exp_2}. As in Experiment 1, we also tested alternative models (see Table \ref{tab:lfe_models_in_exp_2}).

\begin{table}[!ht] 
\centering 
\footnotesize
\caption{Results From Experiment 2}\label{tab:lfe_models_in_exp_2}
% \begin{tabular}{@{\extracolsep{5pt}}lcccccc} 
\resizebox{\textwidth}{!}{ % Start resizing here
\begin{tabular}{@{\extracolsep{5pt}}l>{\centering\arraybackslash}m{1.5cm}>{\centering\arraybackslash}m{1.5cm}>{\centering\arraybackslash}m{1.5cm}>{\centering\arraybackslash}m{1.5cm}>{\centering\arraybackslash}m{2cm}>{\centering\arraybackslash}m{2cm}} 
\\[-1.8ex]\hline 
\hline \\[-1.8ex] 
 & \multicolumn{6}{c}{\textit{Dependent variable: Compensation likelihood}} \\ 
\cline{2-7} 
\\[-1.8ex] & \multicolumn{6}{c}{Model}\\ 
\\[-1.8ex] & 1 & 2 & 3 & 4 & 5 & 6 \\ 
\hline \\[-1.8ex] 
Constant & 3.337$^{***}$ & & & & & \\ 
  & (0.098) &  & & & & \\ 
 Unedited vs. Edited (Dummy variable) & 0.970$^{***}$ & 1.010$^{***}$ & 0.990$^{***}$ & & 1.011$^{***}$ & 1.002$^{***}$ \\ 
  & (0.114) & (0.084) &(0.084) & & (0.122) & (0.115)\\ 
 Unedited vs. More clear (Dummy variable) & & & & 0.987$^{***}$ & & \\ 
  & & & & (0.103) & & \\ 
 Unedited vs. More coherent (Dummy variable) & & & & 1.090$^{***}$ & & \\ 
  & & & & (0.102) & & \\ 
 Unedited vs. More professional (Dummy variable) & & & & 0.894$^{***}$ & & \\ 
  & & & & (0.102) & & \\ 
\hline \\[-1.8ex] 

Condition & All & All & All & All & \shortstack{Focus \\ on \\ Clarity} & \shortstack{Focus \\ on \\ Professionalism} \\ 
Complaint Content (20 Different Contents) Fixed Effect & N & Y & Y & Y & Y & Y \\ 
Participant Fixed Effect & N & Y & Y & Y & Y & Y \\ 
Complaint Presentation Position (1-5) Fixed Effect & N & N & Y & Y & N & N \\ 
Observations & 1,600 & 1,600 & 1,600 & 1,600 & 800 & 800 \\ 
\hline 
\hline \\[-1.8ex] 
\textit{Note.} $^{*}$\textit{p} $<$ .05; $^{**}$\textit{p} $<$ .01; $^{***}$\textit{p} $<$ .001 \\ 
\end{tabular}
}
\end{table}

Experiment 2 replicated the findings from Experiment 1. Specifically, the complaints edited with ChatGPT (Coherent, Clear, and Professional complaints combined) were more likely to receive hypothetical monetary compensation than the unedited complaints (Control complaints only) in a preregistered linear fixed effects model, regardless of whether participants focused on clarity or professionalism of complaints, \textit{b}s > 1.00, \textit{SE}s < 0.12, \textit{t}s > 8.25, \textit{p}s \textless~.001.

\subsection{Experiment S1}
\label{si_subsection:exp_s1}

\paragraph{Method} Experiment S1 was a nearly direct replication of Experiment 1. We recruited 210 participants ($\textit{M}_{age}$ = 44; 49.5\% men, 50\% women, and 0.5\% other) from Prolific. The procedure of Experiment S1 matched that of Experiment 1, with a minor difference: For each of the five complaints, after rating their likelihood of offering hypothetical monetary compensation (i.e., the main dependent measure), participants also rated the complaint on clarity, coherence, and professionalism (in that fixed order). Specifically, they answered the following three questions right after the main dependent measure, ``How clear [coherent / professional] was the complaint above?'' (1 = \textit{Not clear [coherent / professional] at all}, 7 = \textit{Extremely clear [coherent / professional]}). Aside from this difference, all other aspects of the procedure were the same as in Experiment 1.

\paragraph{Results}
We analyzed data exactly as specified in the preregistration (\href{https://aspredicted.org/MM3_17T}{https://aspredicted.org/MM3\_17T}). Specifically, we fit a linear fixed effects model with the likelihood of offering (hypothetical) monetary compensation as the dependent variable and whether or not a given complaint was edited (to be more clear, coherent, or professional) or not as the independent variable, controlling for (1) the fixed effects of each of the 20 complaint contents and (2) the fixed effects of each of the 210 individual participants (see Model 2 in Table \ref{tab:lfe_models_in_exp_s1}).

\begin{table}[!ht] \centering 
\footnotesize
\caption{Results From Experiment S1}\label{tab:lfe_models_in_exp_s1}
% \resizebox{8.0cm}{!}{%
\begin{tabular}{@{\extracolsep{5pt}}lcccc} 
\\[-1.8ex]\hline 
\hline \\[-1.8ex] 
 & \multicolumn{4}{c}{\textit{Dependent variable: Compensation likelihood}} \\ 
\cline{2-5} 
\\[-1.8ex] & \multicolumn{4}{c}{Model} \\ 
\\[-1.8ex] & 1 & 2 & 3 & 4 \\ 
% \\[-1.8ex] & Model 1 & Model 2\\ 
\hline \\[-1.8ex] 
Constant & 2.779$^{***}$ & & & \\ 
  & (0.113) &  & & \\ 
 Unedited vs. Edited (Dummy variable) & 0.737$^{***}$ & 0.743$^{***}$ & 0.742$^{***}$ & \\ 
  & (0.132) & (0.093) &(0.093) & \\ 
  Unedited vs. More clear (Dummy variable) & & & & 0.810$^{***}$\\ 
  & & & & (0.116) \\ 
  Unedited vs. More coherent (Dummy variable) & & & & 0.720$^{***}$\\ 
  & & & & (0.115) \\ 
  Unedited vs. More professional (Dummy variable) & & & & 0.698$^{***}$\\ 
  & & & & (0.116) \\ 
\hline \\[-1.8ex] 

Complaint Content (20 Different Contents) Fixed Effects & N & Y & Y & Y\\ 
Participant Fixed Effects & N & Y & Y & Y\\ 
Complaint Presentation Position (1-5) Fixed Effects & N & N & Y & Y\\ 
Observations & 1,050 & 1,050 & 1,050 & 1,050\\ 
\hline 
\hline \\[-1.8ex] 
% \textit{Note.}  & \multicolumn{4}{r}{$^{*}$p$<$0.05; $^{**}$p$<$0.01; $^{***}$p$<$0.001} \\ 
\textit{Note.} $^{***}$\textit{p} $<$ .001 \\ 
\end{tabular}%
% }
\end{table}

Consistent with our hypothesis (and as mentioned in the main text), the complaints edited with ChatGPT were more likely to receive (hypothetical) monetary compensation than the unedited complaints, \textit{b} = 0.743, \textit{SE} = 0.093, \textit{t} = 7.96, \textit{p} \textless~.001 (Model 2 in Table \ref{tab:lfe_models_in_exp_s1}). Although not preregistered, we fit another linear fixed effects model, which also controlled for the presentation order of complaints (Model 3 in Table \ref{tab:lfe_models_in_exp_s1}). We find the same results, \textit{b} = 0.742, \textit{SE} = 0.093, \textit{t} = 7.94, \textit{p} \textless~.001. Lastly, we examined which type of edits (``more clear'' vs. ``more coherent'' vs. ``more professional'') results in the biggest increase in the likelihood of receiving (hypothetical) monetary compensation by fitting yet another linear fixed effects model (Model 4 in Table \ref{tab:lfe_models_in_exp_s1}), adding the three types of edit as the three independent dummy variables (unedited = 0, each type of edit = 1) and removing the previous binary variable of whether a given complaint was edited or not. We find that the complaints edited to be more clear showed the directionally greatest increase in likelihood of monetary compensation (\textit{b} = 0.810, \textit{SE} = 0.116), as compared with those edited to be more coherent (\textit{b} = 0.720, \textit{SE} = 0.115), or those edited to be more professional (\textit{b} = 0.698, \textit{SE} = 0.116); see Model 4 in Table \ref{tab:lfe_models_in_exp_s1}.

\begin{figure}[!ht]
    \centering
    \includegraphics[width=10.6cm]{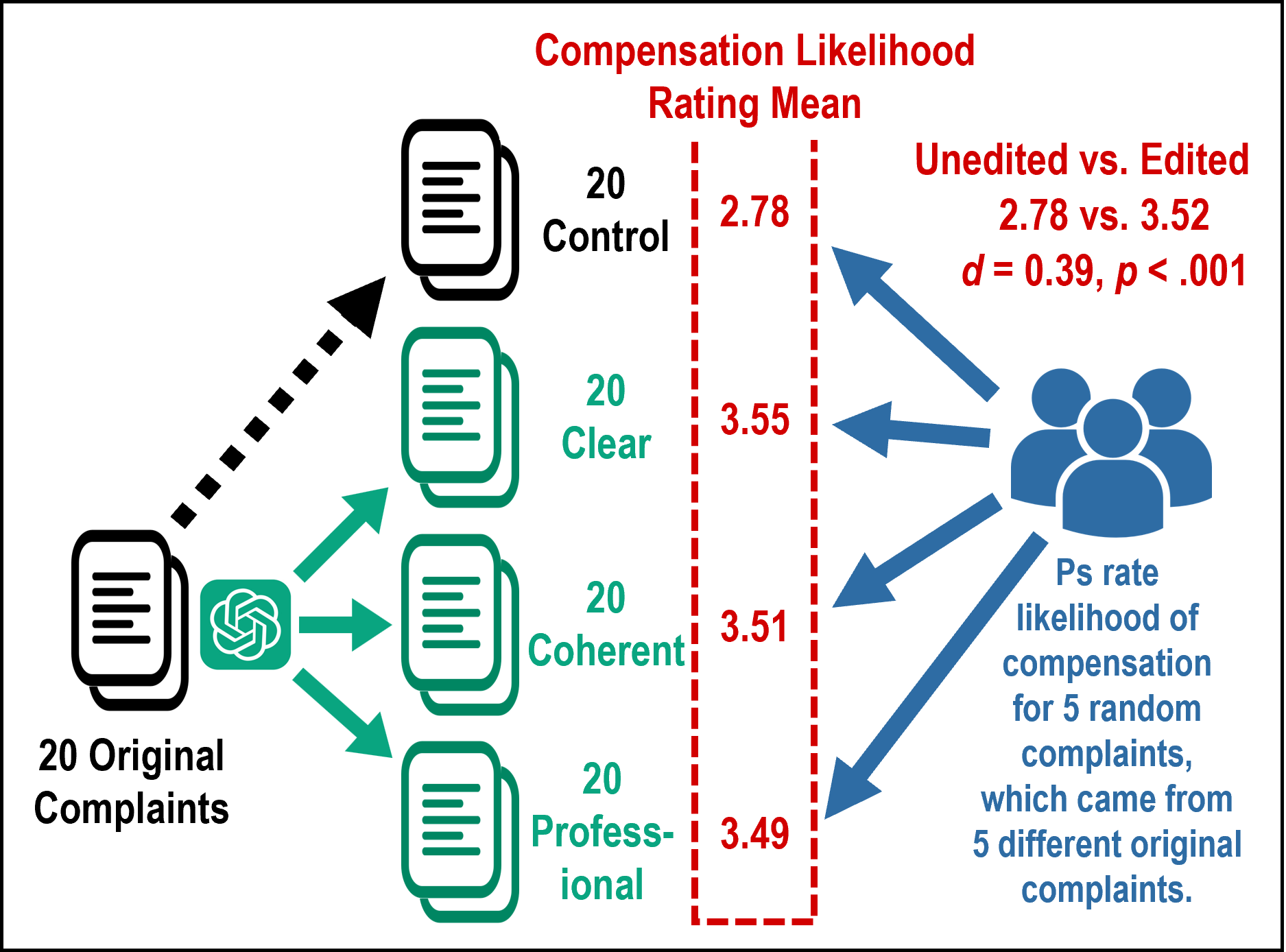}
    \caption{The figure summarizes the procedure and results of Experiment S1. The 20 original complaints were edited with ChatGPT to be more clear, more coherent, or more professional, which resulted in 20 unedited and 60 edited complaints (20 Control, 20 Clear, 20 Coherent, and 20 Professional complaints). Participants in Experiment 1 were presented with a random set of 5 complaints from this pool of 80, viewing one complaint at a time. For each of the 5 complaints, they imagined handling the complaint and first rated the likelihood of offering monetary compensation in response to the complaint, and then evaluated the complaint's clarity, coherence, and professionalism.}
    \label{fig:exp_1_methods_and_results}
    \vspace{-0.15cm}
\end{figure}

\section{Note on the Design of Experiments}
\label{si_subsection:note_on_exp_design}

For the main experiments, we could have chosen to manipulate linguistic features using a study design in which each participant would be randomly assigned to receive one complaint out of the 80 (Alternative Design), rather than a series of multiple complaints (e.g., 5 complaints in a within-subjects design). However, we had a few justifications for opting for our preregistered, within-subjects study designs instead. 

First, presenting multiple complaints as in our design likely led participants to form their own standard for good complaints (which in turn would make them better judges), whereas such a process would be unlikely to occur if participants saw only one complaint out of the 80 (Alternative Design), as is typically done in between-subjects experiments where often a single, rather than multiple, stimulus is presented. By presenting multiple complaints, participants were naturally led to compare complaints to one another, a process through which they likely came to establish their own standards of what makes a good complaint. Such standards then would have made them better judges of complaints, compared to evaluating a single complaint (Alternative Design).

Second, our design more closely matched the real-world setting where evaluators of consumer complaints encounter a wide variety of complaint types and multiple of them, rather than exclusively encountering one specific type of complaint or one specific complaint. Thus, exposing participants to \textit{multiple} complaints of \textit{different} types made our study more ecologically valid compared to exposing them to a single complaint (Alternative Design). In sum, our study design likely enhanced the robustness and real-world applicability of our study's findings.

\end{document}